\documentclass[aps,pra,onecolumn]{revtex4-2}

\usepackage{graphicx}
\usepackage{amsmath,amsfonts,amssymb}
\usepackage{color, xcolor}
\usepackage{subfigure}
\usepackage{physics}
\usepackage{dsfont}
\usepackage{hyperref}
\usepackage{url}
\usepackage{leftidx}
\usepackage{textcomp}
\usepackage{gensymb}
\usepackage[normalem]{ulem}
\usepackage{dsfont}
\usepackage{enumitem}
\usepackage{braket}
\usepackage{pifont}
\usepackage{hyperref}

\definecolor{emerald}{rgb}{0.31, 0.78, 0.47}
\definecolor{blue(ncs)}{rgb}{0.0, 0.53, 0.74}
\definecolor{emerald2}{rgb}{0.25, 0.73, 0.57}
\definecolor{benjoscolour}{rgb}{1.00, 0.50, 0.00}
\definecolor{orchid}{rgb}{0.6, 0.2, 0.8}

\hypersetup{
     colorlinks=true,
     linkcolor=blue,
     filecolor=blue,
     citecolor=emerald,      
     urlcolor =blue(ncs),
}

\newcommand{\e}{\mathrm{e}}
\def\ii{\mathrm{i}}
\newcommand{\pd}{\partial}
\newcommand{\EV}{\mathrm{eV}}

\newcommand{\+}{\, +\,}

\DeclareMathAlphabet{\pazocal}{OMS}{zplm}{m}{n}

\definecolor{emerald}{rgb}{0.31, 0.78, 0.47}
\definecolor{blue(ncs)}{rgb}{0.0, 0.53, 0.74}
\definecolor{emerald2}{rgb}{0.25, 0.73, 0.57}
\definecolor{benjoscolour}{rgb}{1.00, 0.50, 0.00}
\definecolor{orchid}{rgb}{0.6, 0.2, 0.8}

\begin{document}

\author{Alexander V. Balatsky}
\email{balatsky@hotmail.com}
  \affiliation{Nordita, KTH Royal Institute of Technology and Stockholm University, Roslagstullbacken 23,  SE-106 91 Stockholm, Sweden}
  \affiliation{Department of Physics and Institute for Materials Science, University of Connecticut, Storrs, CT 06269, USA}  

\author{Benjo Fraser}
  \email{benjojazz@gmail.com}
  \affiliation{Nordita, KTH Royal Institute of Technology and Stockholm University, Roslagstullbacken 23,  SE-106 91 Stockholm, Sweden}
 
 \author{Henrik S. R{\o}ising}
  \email{henrik.roising@su.se}
  \affiliation{Nordita, KTH Royal Institute of Technology and Stockholm University, Roslagstullbacken 23,  SE-106 91 Stockholm, Sweden}

\title{Dark Sound}
\begin{abstract}
We discuss the axion dark matter (DM) condensate and the consequences the interactions of dark matter would have on the spectrum of collective modes. We find that DM self-interactions change the spectrum of excitations from a quadratic to a linear-like dispersion with velocity $v_s$ which is set by the interactions, but dominated by gravity. For typical DM densities and interactions we find $v_s \sim 10^{-12}c$. This sound-like  mode corresponds to DM density oscillations just like in any other Bose liquid, hence we call it {\em Dark Sound} (DS). The DS mode is well defined and describes stable density oscillations at intermediate length scales $k \geq k_{\text{min}}  \sim 10^{4}\mathrm{lyr^{-1}}$. In the extreme long wavelength limit gravity dominates and leads to Jeans instability of the sound mode at the scale of clump formation $ k \leq k_{\text{min}}$.  We also discuss the possible observable consequences of the DS, including quantized DS modes inside clumps, their characteristic energy, and noise features that might facilitate the observation of DM. 
\end{abstract}

\maketitle

\tableofcontents

%
\section{Introduction}
\label{sec:Introduction}
%
Dark Matter (DM)~\cite{Peebles:2013hla} makes a vivid gravitational imprint on visible matter, but the rest of its nature is yet to be understood. This remains one of the main challenges in modern physics. It makes up over 75\% percent of mass in the universe, but its non-gravitational interactions with visible matter are so weak that it remains hidden, and detection of its signatures is one of the most pressing ongoing experimental challenges~\cite{Feng10}. 

Although DM interacts weakly with the observed matter, one would expect that it is self-interacting. First, there is  gravity: it is assumed that DM exerts gravitational pull on other matter, both baryonic and DM. Second, there are local interaction terms like DM density-density interactions that emerge in the low energy limit of DM action. DM interactions, however small, will have an observable effect on the low energy excitation spectrum of the DM field.  Interactions are known to drastically modify the free particle dispersion: in the non-relativistic free Bose gas with quadratic dispersion, turning on small repulsive interaction modifies the spectrum from quadratic to a linear sound mode $\omega(k) = k^2/2m \rightarrow v_s k $, with a Bogoliubov sound velocity $v_s$ induced by density-density interaction~\cite{pitaevskii2003}. The sound mode in the Bose gas is a reflection of the increased potential energy at higher density (i.e. positive compressibility) that produce the  restoring force. Any Bose liquid possesses Bogoliubov sound modes. Similarly, Fermi liquids exhibit sound modes that emerge as a result of Fermi liquid interactions~\cite{landau1980course}. The existence of Bose and Fermi liquid hydrodynamic sound modes is a universal feature of interacting liquids and does not depend on the relativistic vs. non-relativistic regime. In the long wavelength (IR) limit they are Goldstone modes for spontaneously broken particle number symmetry. 

We thus ask the question about the nature of the low energy excitations of interacting DM. Before we proceed with more detailed discussion, let us summarize our approach and results: 
\begin{itemize} \renewcommand{\labelitemi}{$\star$}
    \item We consider a specific case of Bose statistics in DM: QCD axion DM. We assume the light axion DM is in the regime of strong overlap of wave functions of the axion particles, and that this results in a Bose condensate. 
    \item We find that interactions indeed induce collective DM modes that can be viewed as sound. We call this mode {\it Dark Sound (DS)}. We estimate the typical velocity of the DS modes and find it to be of order $10^{-4}\mathrm{ms}^{-1}$.
    \item In a major departure from the case of repulsive interactions, both the gravitational forces and density-density interactions are attractive in the model. As a result, the Bose liquid is unstable in the IR limit and forms inhomogeneous states, so called {\em clumps}~\cite{Guth:2014hsa}. We find finite size {\em stable} clumps. 
    \item Because of their finite size, the DS modes localized on the clumps have a discrete spectrum. The stable clump supports a single localized sound normal mode. This localized mode will in principle have observable consequences: for example, it will affect the specific heat of the clump. 
    \item We find that coupling of DM to sources like photons has the potential to stabilize clumps and induce a gap in the DS mode. 
\end{itemize}
The typical scale of characteristic clumps, DS sound speed and other quantities are listed in Table~\ref{tab:EstimatesIntro}. 

\vspace{3pt}

\begin{center}
    \ding{167}
\end{center}

\vspace{3pt}

The QCD axion is a hypothetical pseudoscalar boson which was originally theorized to solve the strong CP problem~\cite{PecceiQuinn77, Weinberg78, Wilczek78} in quantum chromodynamics. Such a particle has not yet been detected, which puts strong constraints on its coupling strength to visible matter. For QCD axions the coupling strength is proportional to the axion mass, so that its mass must be very small, with typical expectations in the $10^{-6}$--$ 10^{-3}\EV$ range~\cite{Zyla:2020zbs}. 

We consider axionic DM. In this well-motivated approach, \textit{axions} make up at least part of the `cold dark matter' (CDM) of the $\Lambda$CDM cosmological model~\cite{Preskill:1982cy}.  It has been argued~\cite{Sikivie:2009qn, Arvanitaki:2019rax} that axionic dark matter has experimental signatures distinct from conventional CDM candidates. In particular, because the axion is light it can be quantum mechanically degenerate. Let us assume for argument that the axion density is the mean cold matter density in our galaxy ($\sim 300 \mathrm{MeV}/\mathrm{cm}^3$~\cite{Read_2014}), the mass is $\sim 1 \mu\EV$, and the velocity is the typical galactic velocity ($v\sim 10^{-3}c$~\cite{LawsonEA19}). Then the de Broglie wavelength is $10^3\mathrm{m}$ while the mean separation is $\sim 10^{-5}\mathrm{m}$. Therefore the single particle wavefunctions will overlap considerably. This is a necessary condition for the formation of a macroscopic coherent state: a Bose--Einstein condensate (BEC). This paper takes as its hypothesis that axions form such a condensed phase. We can therefore use classical field theory, as has been argued in e.g.~\cite{Guth:2014hsa}

Significant attention in the field is devoted to the notion that the BEC of bosonic DM is subject to attractive interactions.  If this is the case then a homogeneous ground state is not possible and the BEC state forms clumps with no infinite-range phase coherence. While correct, this point should not preclude one from asking questions like: what is the typical size of the clumps? If this scale is larger than the measurement scale (say table-top experiments) then for all intents and purposes the condensate is homogeneous. We can go about investigating local BEC properties as measured within the clump, such as phase coherence and sound modes. Indeed we find that there are stable clump solutions with the $10^{12}$m scale (see Table~\ref{tab:EstimatesIntro}) that could provide such a region of `local' phase coherence. 

\footnotetext[5]{We will address this question in a separate paper.}

The DS modes we find represent excitations in the fabric of DM. Density fluctuations of DM due to DS modes are widely present in the universe. One immediate consequence of the DS excitations is the presence of density fluctuations and  noise in axion field. For the fluctuations of the axion field away from its mean value we have a finite $\braket{\delta \hat{a}^2({\bf r},t)}, \delta \hat{a}({\bf r},t) = \hat{a}({\bf r},t)- \braket{\hat{a}}$, and the size of these fluctuations will be controlled by the characteristic energy of DM that, presumably, is set by the decoupling energy of dark sector. These DS modes are ubiquitous. The situation resembles the case of black body radiation (BBR), which is an ubiquitous background signal in sky due to photons that have a $3\mathrm{K}$ characteristic energy. Observation of DS and its corresponding DM BBR could open an alternative route to DM detection~\cite{Note5}. 

The outline of this paper is as follows. In Sec.~\ref{sec:AxionInteractions} we review the low energy action for the QCD axion. In Sec.~\ref{sec:ModellingCondensate} we introduce axion bosonic field operators, and develop a BEC coherent state description which is interpreted in Sec.~\ref{sec:meaning}. In Sec.~\ref{sec:condensatesolution} we discuss the collective excitations of a homogeneous BEC state as enabled by tuning the electromagnetic coupling; in Sec.~\ref{sec:sound} we define and discuss the sound mode in such a condensate. In Sec.~\ref{sec:inhomogeneous} we turn to the gravity-dominated inhomogeneous condensate. We discuss the form of the clump solutions in Sec.~\ref{sec:FormClumps} and the sound mode in Sec.~\ref{sec:SoundClumps}, and in Sec.~\ref{sec:NumericalSimulations} we provide results of computer simulations. We give numerical estimates of observables in Sec.~\ref{sec:Estimates}, and discuss experimental signatures in Sec.~\ref{sec:SignalShape} and \ref{sec:HeatCapacity}. Finally, in Sec.~\ref{sec:conclusion} we summarize the results and provide an outlook for possible observable consequences of DS in our search for DM signatures. 

\begin{table}[t!bh]
\centering
\caption{Estimates of key quantities characterizing the dark matter condensate. The dark sound (DS) speed $v_s$ is the typical velocity of the longitudinal dark matter density modes at the characteristic momentum where the modes are stable (Sec.~\ref{sec:AxionBare}); $\Delta$ describes the gap induced by the axion-photon coupling in the dark sound spectrum (Sec.~\ref{sec:AxionPhoton}); $r_0$ is the the characteristic size of the inhomogeneous dark matter clumps (Sec~\ref{sec:FormClumps}); $\mu$ is the chemical potential of the inhomogeneous condensate (Sec.~\ref{sec:FormClumps}); $\omega_{\mathrm{GS}}$ is the normal mode ground state energy for the quantized dark sound spectrum in a clump (Sec.~\ref{sec:SoundClumps}).}
\label{tab:EstimatesIntro}
\begin{tabular}{ p{4.5cm}  p{8cm} }
	\hline \hline
     HOMOGENEOUS &  \\ \hline
     Sound speed & $v_s\, =\, \left(\frac{10^{-6}\EV}{m} \right)^{1/2}\left(\frac{\rho_{\text{DM}}}{\rho}\right)^{1/4} 10^{-12}c$ \\
     Photon gap & $\Delta \sim \left(\frac{g_{a\gamma \gamma}}{10^{-15}\mathrm{GeV}}\right)\left(\frac{\lvert \boldsymbol{E} \cdot \boldsymbol{B} \rvert}{10\mathrm{T} \cdot  10^6\mathrm{V}/\mathrm{m} }\right)\left(\frac{\rho_{\text{DM}}}{\rho}\right)^{1/2} 10^{-18} \mathrm{eV}$ \\ \hline
     CLUMP &  \\ \hline
     Clump size & $r_0\, \sim\, \left(\frac{10^{-6}\mathrm{eV}}{m} \right)^{1/2}\, \left(\frac{\rho_{\text{DM}}}{\rho}\right)^{1/4}\, 10^{12}\mathrm{m}$ \\
     Clump chemical potential& $\mu\sim \left(\frac{\rho}{\rho_{\text{DM}}}\right)^{1/2}10^{-30}\mathrm{eV}$ \\
     Normal mode energy& $\omega_{\mathrm{GS}}=b\mu$\quad universal factor $b\simeq 0.54$ \\
	\hline \hline
\end{tabular}
\end{table}

%
\section{Axion model}
\label{sec:model}
%
%
\subsection{Axion interactions}
\label{sec:AxionInteractions}
In this paper we treat the case of the QCD axion (for reviews see Refs.~\cite{Kim:1986ax,Marsh:2015xka}), although more general axion-like particles can be considered~\cite{Marsh:2015xka}. Axions can couple to both themselves and to standard model matter, as well as through gravity. The QCD axion is defined by its coupling to gluons, of the form $a G_{\mu\nu}\tilde{G}^{\mu\nu}$. We work with the theory at energies below the mass of the pions, where the heavy QCD degrees of freedom are integrated out. The action for the axion is
\begin{align}
    \label{eq:relaction}
    \mathcal{L}\, =\, \sqrt{-g}\left[\frac{1}{16\pi G}\mathcal{R}\, -\, \frac{1}{2}g^{\mu\nu}\pd_{\mu} a\pd_{\nu} a\, -\, V(a)\+ g_{a\gamma\gamma}\frac{e^2}{16\pi^2} \frac{a}{f_a} F_{\mu\nu}\tilde{F}^{\mu\nu}\, -\, g_{aee}\frac{m_e}{f_a}  \ii\, a\bar{\psi}_e\gamma^5 \psi_e\, + \dots \right],
\end{align}
where $G$ is Newton's gravitational constant, $g_{\mu\nu}$ is the spacetime metric, $g$ its determinant, and $\mathcal{R}$ is the Ricci scalar of $g_{\mu\nu}$. Above, `$\dots$' refers to other matter couplings~\cite{TrickleEA20}. The non-gravitational couplings are all small because they are controlled by inverse powers of the axion energy scale $f_a$, which is constrained by experiment to be very large: $f_a\gtrsim 10^7\mathrm{GeV}$ \cite{Zyla:2020zbs}. Throughout this paper we will use as a benchmark a QCD axion with the cosmologically favoured mass $\sim 10^{-5}\mathrm{eV}$ \cite{Raffelt:2006cw}, corresponding to $f_a\sim 10^{12}\mathrm{GeV}$. 

Axions self-interact gravitationally via the standard minimal coupling of general relativity \cite{carroll2003spacetime}. The potential $V(a)$ can be found using chiral perturbation theory~\cite{diCortona:2015ldu}
\begin{align}
\label{eq:axionpotential}
V(a)\, =\, m_{\pi}^2f_{\pi}^2\sqrt{1-\frac{4m_um_d}{(m_u+m_d)^2}\sin^2 \left( \frac{a}{2f_a} \right)}\, =\, \mathrm{constant}\, + \frac{1}{2}m^2\, a^2 + \frac{\lambda}{4!}a^4\, + \dots,
\end{align}
where $m_u$ and $m_d$ are the up and down quark masses. This expression is accurate to about $1\%$ of the exact value \cite{diCortona:2015ldu}, more accurate than the commonly quoted cosine potential. A useful quantity is the topological susceptibility $\chi$, which is identically equal to $\chi = m^2 f_a^2$~\cite{diCortona:2015ldu}, and is set by characteristic QCD energy scales (we define the scale $\Lambda_{\mathrm{QCD}}\equiv\chi^{1/4} \simeq 76\mathrm{MeV}$). Working out the expansion in Eq.~\eqref{eq:axionpotential} we find that
\begin{align}
    \label{eq:quarticcoupling}
    \lambda\, =\, -\frac{1}{2}\left( \frac{m_u-m_d}{m_u+m_d} \right)^2 \frac{\Lambda_{\mathrm{QCD}}^4}{f_a^4}\quad . 
\end{align}
Therefore the leading interaction term is quartic with \textit{negative} sign, corresponding to an \textit{attractive} delta function potential self-interaction. 

Coupling to other matter takes different forms depending on the precise model under consideration. We would like to draw particular attention to the photon and electron couplings written in \eqref{eq:relaction}. For the two most prominent models (the KSVZ and DFSZ axions)~\cite{Marsh:2015xka}, the coupling constants are $g_{a\gamma\gamma}=-1.92$, $0.75$ and $g_{aee}=0$, $\frac{2\sin^2\beta}{3}$ respectively, where $\tan\beta\equiv \braket{H_u}/\braket{H_d}$ is the ratio of the two Higgs expectation values in the DFSZ model. We have used the commonly quoted quark mass ratio of $z\equiv m_u/m_d = 0.56$~\cite{Leutwyler:1996qg, Raffelt:2006cw}.

%
\subsection{The axion condensate}
\label{sec:ModellingCondensate}
%
The axion field $a$ in the Lagrangian of Eq.~\eqref{eq:relaction} is a real field, so in a second-quantized theory theory it corresponds to the real part of the (Heisenberg picture) creation operator. An apt analogy is the displacement of a harmonic oscillator expressed in terms of creation and annihilation operators: $\hat{\bf{x}} = \hat{b} + \hat{b}^{\dag}$. In exact analogy with that case we promote $a$ to an operator $\hat{a}$ which we decompose in terms of complex creation and annihilation operators $\hat{\psi}$, $\hat{\psi}^{\dagger}$:
\begin{align}
\label{eq:quantumfield}
\hat{a}\, \equiv \, \frac{1}{\sqrt{2m}}\left[ \hat{\psi}(t,\mathbf{x})\, \e^{-\mathrm{i}(m+\mu)t} + \hat{\psi}^{\dagger}(t,\mathbf{x})\, \e^{\mathrm{i}(m+\mu)t} \right].
\end{align}
Notice that, in constrast with standard expressions, we have allowed for a non-zero chemical potential $\mu$ in \eqref{eq:quantumfield}: the axion field can be oscillating at a frequency different from its mass because of interactions in a finite density state. 

One often introduces the amplitude and the phase of $\hat{\psi} = \sqrt{\hat{n}} \exp\big(i \hat{\theta}\big)$ which has both a magnitude \textit{and} a phase. Therefore we can think of there being two variables $\hat{n}, \hat{\theta}$ at each point in space. 
In the case where we have a large occupation number ($\langle \hat{\psi}^{\dag}\hat{\psi} \rangle \gg1$) it is possible that they form a coherent (Glauber) state, characterized by an expectation value for the \textit{complex} field operator $\hat{\psi}$.  We assume that the axion is in such a coherent state, which we refer to as the `condensate', and that the expectation value $\psi\equiv\langle \hat{\psi} \rangle$ behaves like a classical field. Note that this is a non-trivial assumption: see for example \cite{Hertzberg:2016tal} and recent work \cite{Kopp:2021ltb}. Given this assumption, we can remove the hats in Eq.~\eqref{eq:quantumfield} and treat the $\psi$s as commuting numbers. 

We take as our starting point the non-relativistic action for the axion condensate coupled to photons of frequency $\omega_{\gamma}$ through the coupling in Eq.~\eqref{eq:relaction}, which is derived in Appendix~\ref{sec:NonrelativisticAction} following standard approximations. The non-relativistic limit can be justified if it is assumed the axions have galactic velocities, $v/c\sim 10^{-3}$ \cite{LawsonEA19}. In order for the photon interaction term in Eq.~\eqref{relaction} to survive this limit, $\omega_{\gamma}$ must be of the order of the Compton frequency $m$. To keep expressions concise we package the photon coupling into a coefficient $g$, which is essentially the axion-photon coupling multiplied by the amplitude of the electromagnetic wave background. We define
\begin{align}
    -g_{a\gamma\gamma}\frac{e^2}{4\pi^2} \frac{a}{f_a} \mathbf{E}\cdot\mathbf{B}\equiv g\, \e^{\mathrm{i}\omega_{\gamma}t}\, a + \mathrm{c.c.}
\end{align}
where `c.c.' means the complex conjugate. In Sec.~\ref{sec:condensatesolution} we will limit ourselves to photons such that $\omega_{\gamma} - m - \mu=0$, so that the coupling is basically $g(a+a^*)$, but for now we keep $\omega_{\gamma}$ general. 

While we have taken the photon coupling as an example, many different fields have the same form of the coupling to the axion. For example, see the electron coupling in the action \eqref{eq:relaction}. We have also not included the kinetic terms for the photons, so that we are are neglecting backreaction. Inclusion of these terms leads to interesting phenomena including mixing of the photon with DS modes. 

The action once we have taken all our limits (see Appendix~\ref{sec:NonrelativisticAction} for details) reads
\begin{align}
\label{nonrelaction}
\mathcal{L}\, =\, &\frac{1}{2}(\ii\, \psi^*\partial_t\psi + \text{c.c.})\, +\, \mu \lvert \psi \rvert^2\, -\, \frac{1}{2m}\lvert \nabla \psi\rvert^2 - \frac{\lambda}{16 m^2}\lvert \psi \rvert^4\nonumber\\
&\, +\, \big(\frac{g}{\sqrt{2m}}\e^{\mathrm{i}(\omega_{\gamma} - m - \mu)t} \psi + \text{c.c.}\big) + \frac{1}{2} G m^2 \int \dd^3 x' \frac{\lvert \psi({\bf x})-\bar{\psi} \rvert^2 \lvert \psi({\bf x}')-\bar{\psi}\rvert^2}{\lvert {\bf x}-{\bf x}' \rvert} .
\end{align}
Here the zero mode $\bar{\psi}$ of $\psi$ has been subtracted off in the gravitational term (see the appendix for details). The Lagrangian of Eq.~\eqref{nonrelaction} has the corresponding equation of motion
\begin{align}
    \label{eq:eom}
    \ii\, \partial_t\psi\, =\, -\frac{1}{2m}\nabla^2 \psi\, -\, \mu \psi \+ \frac{\lambda}{8 m^2}\lvert \psi \rvert^2\psi + \Phi_{\mathrm{N}}(\psi)\psi - \frac{g}{\sqrt{2m}}\e^{\mathrm{i}(\omega_{\gamma} - m - \mu)t},
\end{align}
where $\Phi_{\mathrm{N}}(\psi)$ is the Newtonian potential, defined as
\begin{align}
    \Phi_{\mathrm{N}}(\psi)\, \equiv\, -G m^2 \int \dd^3 x' \frac{\lvert \psi({\bf x}')-\bar{\psi} \rvert^2}{\lvert {\bf x}-{\bf x}' \rvert}.
\end{align}
We can alternatively treat $\Phi_{\mathrm{N}}$ as a separate variable satisfying a separate Poisson equation $\nabla^2\Phi_{\mathrm{N}}=4\pi Gm^2\lvert\psi-\bar{\psi}\rvert^2$.

We decompose $\psi$ into its magnitude and phase $\psi\equiv \sqrt{n}\e^{\ii \theta}$ so that the action~\eqref{nonrelaction} becomes
\begin{align}
\label{eq:thetaphilagrangian}
\mathcal{L}\, =\, &-n\,\dot{\theta} -  \frac{1}{2m}n(\nabla \theta)^2 + n\mu -  \frac{1}{2m}\frac{(\nabla n)^2}{4n}  - \frac{\lambda}{16 m^2}n^2 + \frac{1}{2} G m^2 \int \dd^3 x' \frac{n({\bf x})n({\bf x}')}{\lvert {\bf x}- {\bf x}' \rvert}\nonumber\\
 &+ \frac{g}{\sqrt{2m}} \sqrt{n} \left(\e^{\mathrm{i}(\omega_{\gamma} - m - \mu)t}\e^{\ii\theta} + \e^{-\mathrm{i}(\omega_{\gamma} - m - \mu)t}\e^{-\ii\theta}\right).
\end{align}
The equations of motion in terms of $n$ and $\theta$ are
\begin{align}
    \label{eq:eomntheta}
    \dot{n}\, &=\, -\frac{1}{m}\nabla\cdot (n\nabla\theta)-\ii\frac{g}{\sqrt{2m}} \sqrt{n} \left(\e^{\mathrm{i}(\omega_{\gamma} - m - \mu)t}\e^{\ii\theta} - \e^{-\mathrm{i}(\omega_{\gamma} - m - \mu)t}\e^{-\ii\theta}\right),\nonumber\\
    \dot{\theta}\, &=\, \frac{1}{2m}\frac{\nabla^2\sqrt{n}}{\sqrt{n}} - \frac{1}{2m}\nabla^2\theta - \frac{\lambda}{8m^2} n - G m^2 \int\dd^3 x'\frac{n({\bf x}')}{\lvert{\bf x}-{\bf x}' \rvert}
    +\frac{1}{2}\frac{g}{\sqrt{2m}} \frac{1}{\sqrt{n}} \left(\e^{\mathrm{i}(\omega_{\gamma} - m - \mu)t}\e^{\ii\theta} + \e^{-\mathrm{i}(\omega_{\gamma} - m - \mu)t}\e^{-\ii\theta}\right).
\end{align}
For $g=0$ the action of Eq.~\eqref{nonrelaction} has a $U(1)$ shift symmetry in $\theta$, corresponding to particle number conservation, which is spontaneously broken in the finite density solutions we consider. The canonically conjugate charge is simply the number density $n$. Sound modes are the Goldstone modes of this symmetry breaking: the fluctuations of the $\theta$ field. A non-zero $g$ explicitly breaks $U(1)$ because the source photon field can create particles, and correspondingly gives a potential energy term for $\theta$. The Lagrangian of Eq.~\eqref{eq:thetaphilagrangian} will be the starting point of our analysis of the collective modes, i.e.~DS modes of the axion condensate. 
\subsection{The meaning of the condensed phase}
\label{sec:meaning}
There has been some debate and controversy over the idea of axion condensates, and `Bose--Einstein' condensation in particular~\cite{Sikivie:2009qn,Guth:2014hsa,Marsh:2015xka,Hertzberg:2016tal}. In this paper, by `condensate' we mean only the following: the quantum axion field is in a coherent (Glauber) quantum state with expectation value for the field $\hat{\psi}$: $\braket{\hat{\psi}}\equiv \psi$, such that $\psi$ obeys the classical equations of motion. We use the phrases `condensate' and `Bose--Einstein condensate' freely and interchangeably in this paper. 

$\psi$ carries a phase, as in any macroscopic state, and it is possible for this phase to be correlated over large spatial scales: in particular, in can be larger than the scale of an experiment. The correlation of the phase will be determined by the exact dynamics of the axion~\cite{Guth:2014hsa}: this is investigated in Sec.~\ref{sec:inhomogeneous}. In particular, there is no necessary implication that the axion is `thermalized'. Such phases are ubiquitous in a variety of macroscopic quantum phenomena~\cite{Annett:730995}, and we have no reason to expect the axion to be any different. 

Whether the axion is actually in such a condensed phase is an open question. Its large occupation number does not automatically imply classical behaviour (it is a \textit{necessary} but not \textit{sufficient} property).

\section{Homogeneous condensate}
\label{sec:Homogeneous}
%
\subsection{Condensation solution}
\label{sec:condensatesolution}
%
In this section we consider the case of a homogeneous saddle point of the action. It is instructive to see the emergence of collective DS modes in this simplest case. We find  indeed that one can have a saddle point  solution  in the presence of  external EM fields present with fine tuning of parameters. This result itself is interesting as it  suggests DM can have stable homogeneous `islands' where the gravitational pull is stabilized by the EM fields. 
For a general case the gravitational attraction dominates at extreme IR scales producing clumps of DM, the case we will consider in next sections. 

The equations of motion \eqref{eq:eomntheta} are solved by the stationary `condensate' configuration
\begin{align}
\label{condensate}
 n = n_0 \equiv \mathrm{constant} \qquad \text{and} \qquad  \theta=0, \pi,
\end{align}
which is a solution only when the following two equations hold:
\begin{align}
    &\omega_{\gamma} = m + \mu\quad &\text{Monochromatic photons}\label{eq:monochromatic}\\
    &\frac{\lambda}{8 m^2} n_0^2 - \frac{g}{\sqrt{2m}} \sqrt{n_0} - \mu n_0 = 0\quad &\text{Equilibrium condition}\label{eq:equilibrium}
\end{align}

The interpretation of these equations is as follows. We imagine that the axion field coexists with a population of photons. To study the simplest possible situation, we restrict to photons of a single frequency of Eq.~\eqref{eq:monochromatic}. The axion-photon coupling will induce particle exchange between the axion and photon populations, and we suppose that sufficient time has elapsed to allow them to come into equilibrium locally. Eq.~\eqref{eq:equilibrium} is the condition for this equilibrium. 

The conditions of Eqs.~\eqref{eq:monochromatic} and \eqref{eq:equilibrium} fix both $\omega_{\gamma}$ and $g$. This is not the most general situation. Photons of other frequencies will be involved, and induce non-linear oscillations. But the resulting system does not have a stationary solution, and so perturbations will not fall into energy eigenstates. This is important to consider in the future, but falls outside the scope of our study. 

Perturbing around the solution of Eq.~\eqref{condensate} to linear order gives, up to constants and total derivatives
\begin{align}
\label{linaction}
\mathcal{L}\, =\, -\delta n\, \dot{\delta\theta}\, - \frac{n_0}{2m}(\nabla \delta\theta)^2 -g\sqrt{n_0}\delta\theta^2  +\, \delta n \left( \frac{1}{2m}\frac{1}{4n}\nabla^2 -\frac{g}{\sqrt{2m}} n_0^{-3/2} - \frac{\lambda}{16 m^2} - 2\pi G m^2 \nabla^{-2} \right) \delta n.
\end{align}
This is the coupled Lagrangian of $n$ and $\theta$ which governs the sound dynamics, whose consequences we shall examine in the following sections. We are using the notation $\nabla^{-2}f(\mathbf{x})$ as shorthand to denote the integral operator $-\frac{1}{4\pi}\int \dd^3 x' \frac{f(\mathbf{x}')}{\lvert \mathbf{x}-\mathbf{x}'\rvert}$ appearing in the Newtonian interaction. 
\begin{figure}[t!bh]
\centering
\subfigure[]{\includegraphics[width=0.45\linewidth]{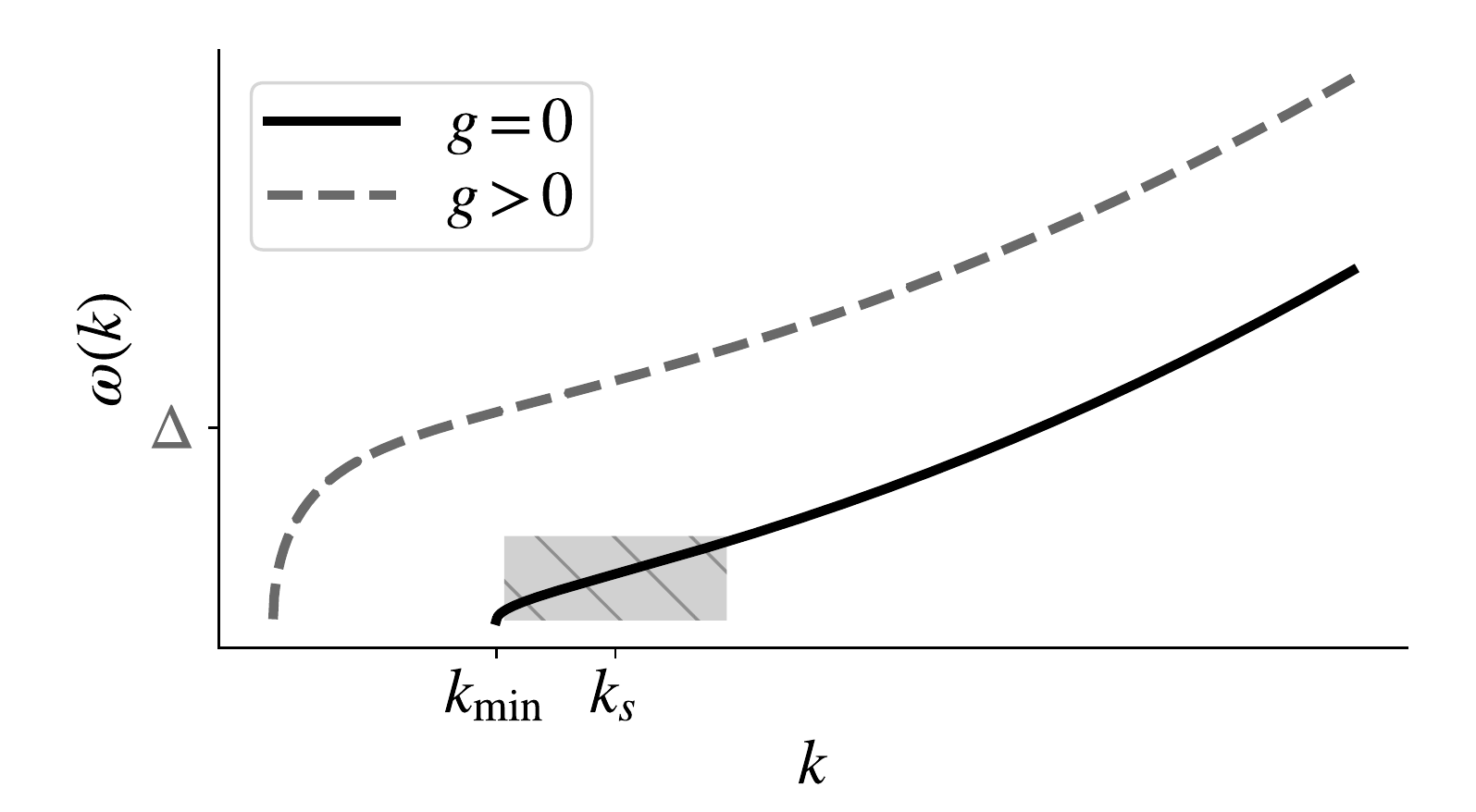} } \quad
\subfigure[]{\includegraphics[width=0.45\linewidth]{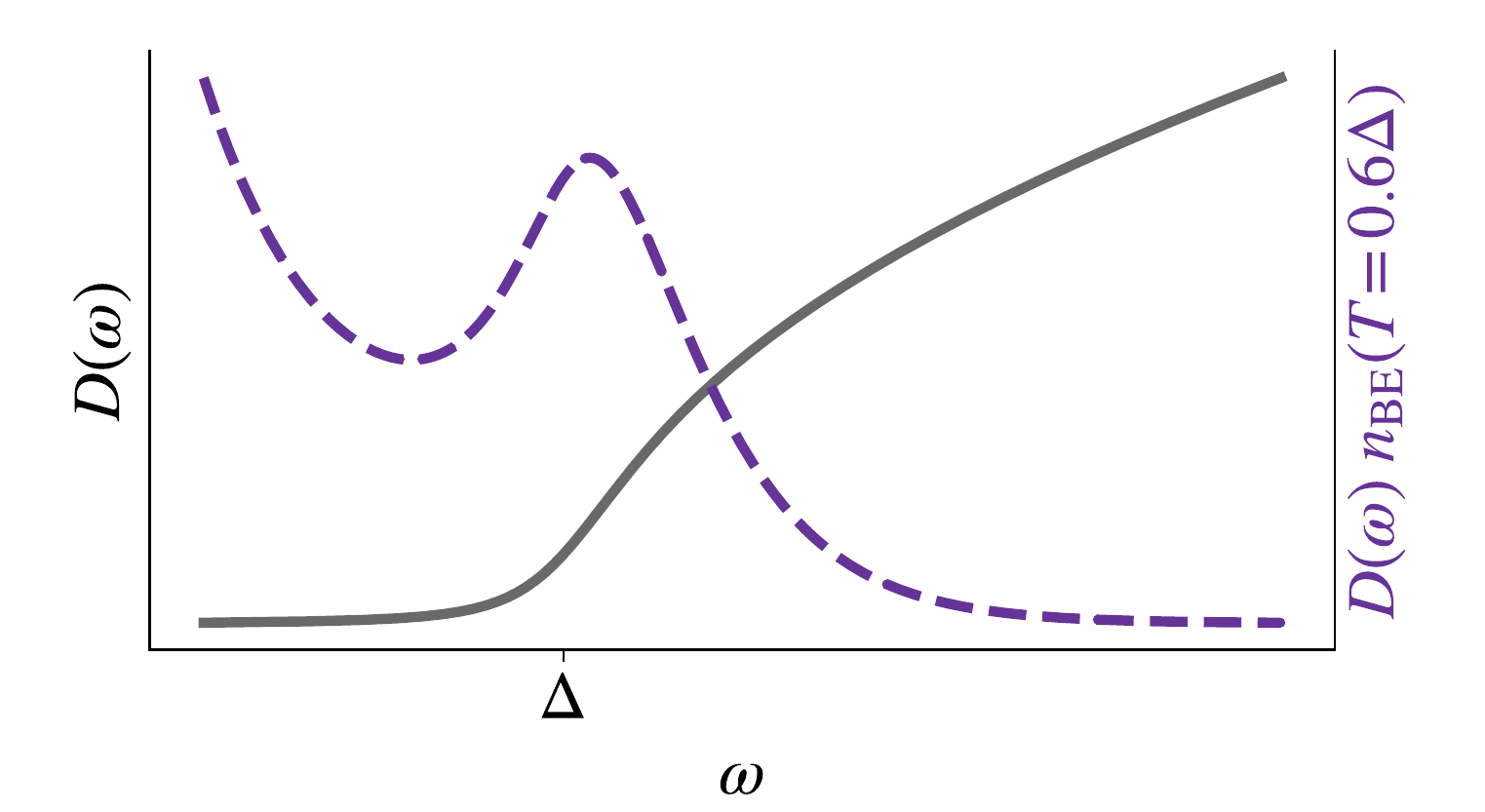} } \quad
\caption{(a) The dispersion relation of Eq.~\eqref{eq:dispersion} as a function of momentum $k$ about the homogeneous condensate. On the axes we have indicated the gap $\Delta$ from Eq.~\eqref{eq:Delta}, the momentum $k_{\text{min}}$ from Eq.~\eqref{eq:kmin} at which the instability occurs (for $g = 0$), and the inflection point $k_{s}$ of $\omega(k)$ which defines the sound speed $v_s = \partial \omega / \partial k \lvert_{k = k_s}$ from Eq.~\eqref{eq:soundspeed}. The grey hatched region marks the $k$ range in which the dispersion is approximately linear (for $g = 0$), see Appendix~\ref{sec:soundappendix} for details. (b) The density of states, $D(\omega) = 4\pi k^2\left(\dd \omega/\dd k\right)^{-1}$ (grey, full line), about the homogeneous condensate and the occupancy at finite temperature (magenta, dashed line), both for the case of $g>0$ in panel (a). Here, $n_{\text{BE}}(T) = (\exp(-\omega/T)-1)^{-1}$ is the Bose--Einstein distribution.}
    \label{fig:Dispersion}
\end{figure}
In order to isolate the dynamics of $\theta$ we integrate out $\delta n$. Solving for $\delta n$ using its equation of motion we find the effective action
\begin{align}
\label{effaction}
\mathcal{L}(\theta)\, =\, -\frac{1}{4} \dot{\delta\theta}\left( \frac{1}{2m}\frac{1}{4n}\nabla^2 -\frac{g}{\sqrt{2m}}\frac{1}{n_0^{3/2}} - \frac{\lambda}{16 m^2} - 2\pi G m^2 \nabla^{-2} \right)^{-1}\, \dot{\delta\theta} - \frac{n_0}{2m}(\nabla \delta\theta)^2 -g\sqrt{n_0}\delta\theta^2,
\end{align}
so we have a wave equation but with a non-standard coefficient for the time derivative. 

The dispersion relation is seen by using a plane wave ansatz
\begin{align}
\label{eq:dispersion}
\boxed{
\omega^2\, =\, \left(\frac{k^2}{2m}+\frac{g}{\sqrt{2 m n_0}}\right)\left(\frac{k^2}{2m}+\frac{g}{\sqrt{2 m n_0}}+\frac{\lambda}{4 m^2} n_0-\frac{8\pi G m^2}{k^2}n_0\right)
}\quad \parbox{12em}{Dark sound,\\ homogeneous case}
\end{align}
This describes the DS modes. It is our key equation in the analysis of perturbations of the homogeneous condensate, and is plotted in Fig.~\ref{fig:Dispersion} (a), with the corresponding density of states plotted in Fig.~\ref{fig:Dispersion} (a). It extends the same equation in for example Ref.~\cite{Guth:2014hsa} by the inclusion of the external coupling $g$. In the next section we discuss properties of the sound. 

%
\subsection{The dark sound mode}
\label{sec:sound}
%
We divide the DS modes of Eq.~\eqref{eq:dispersion} into stable ($\omega^2>0$) and unstable ($\omega^2<0$) modes. The unstable modes will be studied starting in Sec.~\ref{sec:inhomogeneous}. The stable modes will give rise to DS. We first deal with the sourceless axion field, then consider coupling to photons. 

\subsubsection{The axion on its own ($g=0$)}
\label{sec:AxionBare}
Let us first examine the dynamics when external sources are absent ($g=0$). The dispersion relation of Eq.~\eqref{eq:dispersion} becomes
\begin{align}
\label{eq:dispersiong0}
\omega^2\, =\, \frac{k^2}{2m} \left(\frac{k^2}{2m}+\frac{\lambda}{4 m^2} n_0-\frac{8\pi G m^2}{k^2}n_0\right).
\end{align}
In Fig.~\ref{fig:Dispersion} (a) we plot the dispersion and its derivative $\partial\omega/\partial k$ as a function of $k$. 

There is a momentum $k_{\min}$ that determines the stability. High $k$ modes with $k>k_{\min}$ are stable, low $k$ modes with $k<k_{\min}$ are unstable. There are two forces that drive the instability: a gravity ($G$) term and a self interaction ($\lambda$) term. In the case of gravity this is the well-known Jeans instability. To understand the $\lambda$ term, we first ignore the $G$ term in Eq.~\eqref{eq:dispersiong0}, which makes the dispersion relation equivalent to that of Bogoliubov quasiparticles in the weakly interacting Bose gas~\cite{pitaevskii2003}. In the Bogoliubov case $\lambda>0$ (repulsive interactions) $\omega$ is linear in $k$ near $k=0$. This describes the well-known longitudinal density oscillations, i.e.~sound modes, seen in any Bose liquid. For the axion $\lambda<0$ (attractive interactions) and so there is an instability at small $k$ (long wavelengths) which cuts off the dispersion before the linear regime is reached. 

Thus the homogeneous condensate is unstable in the far infrared (IR) limit. However, our key point is: 
\begin{align}
\boxed{\text{For $k>k_{\min}$ there are collective excitations of the condensate which we can call Dark Sound (DS).}}
\end{align}
We now make this precise. The dispersion $\omega(k)$ is quadratic for $k\gg k_{\min}$ (free particles), and unstable for $k< k_{\min}$, as depicted in Fig.~\ref{fig:Dispersion} (a). In the middle of these two regimes we find an inflection point $k=k_s$ where the second derivative vanishes and the dispersion is approximately linear. Near this point the oscillations are stable, but the dispersion relation is also significantly modified from the free quadratic behaviour. Therefore, they are collective sound modes. 

We define the \textit{sound speed} $v_s$ to be the value of $\partial\omega/\partial k$ at $k=k_s$. When gravity dominates we have
\begin{align}
    \label{eq:soundspeed}
    \boxed{
    \left(v_s\right)_{\mathrm{gravity}}\, \equiv\, \left(\frac{\partial\omega}{\partial k}\right)_{k=k_s}\, =\, 3\left(\frac{4\pi}{3} \right)^{1/4} \left(\frac{G \rho}{m^2}\right)^{1/4}
    }\quad \text{Sound speed}
\end{align}
where we have written $\rho\equiv m n_0$ to facilitate comparison with the dark matter density. Note that the sound mode arises entirely from interactions: Eq.~\eqref{eq:soundspeed} vanishes if we send $G\rightarrow 0$.

In situations where the self-interaction is dominant, our arguments still go through, but the sound mode will have a different expression:
\begin{align}
    \left(v_s\right)_{\lambda}\, =\, \left(\frac{|\lambda| \rho}{m^4}\right)^{1/2}.
\end{align}

In appendix \ref{sec:soundappendix} we go into more detail about the momentum regime corresponding to the DS collective modes. Observables related to DS are discussed in Sec.~\ref{sec:observables}.

\subsubsection{The axion coupled to photons ($g\neq 0$)}
\label{sec:AxionPhoton}
When we add an external coupling we must consider the full dispersion relation in Eq.~\eqref{eq:dispersion}. The analysis can be divided into two cases, according to the sign of $g$.

If $g\geq0$, much of the previous section still goes through: there is an instability for $k<k_{\min}$, but the expression for $k_{\min}$ must be modified. It is now the zero of the second bracket in Eq.~\eqref{eq:dispersion}:
\begin{align}
    \frac{k_{\min}^2}{2m}+\frac{g}{\sqrt{2 m n_0}}+\frac{\lambda}{4 m^2} n_0-\frac{8\pi G m^2}{k_{\min}^2}n_0\, =\, 0.
    \label{eq:kminEq}
\end{align}
The $g\geq0$ contribution always pushes $k_{\min}$ to lower values, stabilizing the dynamics for more of $k$-space. To examine the IR behaviour, we expand Eq.~\eqref{eq:dispersion} in powers of $k$:
\begin{align}
    \label{eq:expandeddispersion}
    \omega^2\, =\, -&\frac{g}{\sqrt{2 m n_0}}\frac{8\pi G m^2 n_0}{k^2} + \Delta^2\, +\, \frac{k^2}{2\alpha} + \frac{k^4}{(2m)^2} + \dots,
\end{align}
where we have defined
\begin{align}
\Delta^2\, &\equiv\, \frac{g}{\sqrt{2 m n_0}}\left[ \frac{g}{\sqrt{2 m n_0}}+\frac{\lambda n_0}{4 m^2} - \frac{1}{2m} 8\pi G m^2 n_0\right] \label{eq:Delta}, \\
\alpha\, &\equiv\, \left( \frac{g}{2\sqrt{2 m n_0}}+\frac{\lambda n_0}{4 m^2} \right)^{-1}\, m. 
\end{align}
When $g$ is big enough, $\Delta^2$ is positive. It creates an approximate \textit{gap} for the sound modes, as shown in Fig.~\ref{fig:Dispersion} (a):
\begin{align}
\boxed{\text{The $a\,\mathbf{E}\cdot\mathbf{B}$ coupling to photons creates a gap $\Delta$ in the Dark Sound spectrum. }}
\end{align}

For $g<0$, $\omega$ is real near $k=0$, as evident from the expansion in Eq.~\eqref{eq:expandeddispersion}, but $\omega$ is imaginary within a band of $k$ values: $k^2 \in ( \lvert g \rvert \sqrt{2m/n_0},~k_{\text{min}}^2)$, where $k_{\text{min}}$ is given by Eq.~\eqref{eq:kminEq}. Thus the modes are stable in the far infrared, but have an instability around a specific length scale. This is suggestive of a structure that could emerge with this size when the full Maxwell action is coupled to the axion. We intend to address this in future work. 
%
\section{Clumpy dark matter}
\label{sec:clumpydm}
%
We expect the unstable modes of Eq.~\eqref{eq:dispersion} to evolve into a non-linear regime and produce localized dynamics on the length scale $\sim 1/k_{\mathrm{min}}$: the endpoint of the evolution should be some form of localized \textit{clump}. We investigate the form of the clumps in Sec.~\ref{sec:FormClumps}, then we will investigate the collective modes about this `true' vacuum of the axion field, as opposed the homogeneous solution of Sec.~\ref{sec:Homogeneous}, but first simply consider the instability scale in more detail. 
%
\subsection{Gravity dominates at galactic densities}
\label{sec:inhomogeneous}
%
Let us estimate $k_{\min}$ for the QCD axion, with Eq.~\eqref{eq:quarticcoupling} for the quartic coupling; $k_{\min}$ solves the equation
\begin{equation}
    k_{\mathrm{min}}^4 -\frac{1}{4}\left( \frac{m_u-m_d}{m_u+m_d} \right)^2\frac{\rho\, m^2 }{\Lambda_{\mathrm{QCD}}^4} k_{\mathrm{min}}^2 -8\pi G\rho m^2 = 0.
    \label{eq:kmin}
\end{equation}
The solution interpolates as a function of mass between two limiting behaviours, corresponding to two mass regimes:
\begin{align}
\label{masses}
k_{\min}\sim
\begin{cases}
(G m^2 \rho)^{1/4} & m\ll \Lambda_{\text{QCD}}^4\left( G/\rho \right)^{1/2} \quad \text{gravity regime}\\
m\sqrt{\rho} / \Lambda_{\text{QCD}}^2 & m \gg \Lambda_{\text{QCD}}^4\left(G/\rho\right)^{1/2} \quad \text{quartic coupling regime}\\
\end{cases}
\end{align}
We imagine starting at high $k$ (small spatial scale) perturbations, which are stable. Then we gradually move to lower $k$, and observe the first value, $k=k_{\text{min}}$, for which the evolution becomes unstable. This will be dictated by gravity in the low mass regime of Eq.~\eqref{masses}, or the $\lambda$ term in the high mass regime. Said otherwise: whichever of the two interactions produces the smallest clumps will dictate the spatial structure of the condensate. 

Taking galactic DM density ($\rho \sim 300\mathrm{MeV}\,\mathrm{cm}^{-3}$) we find that $\Lambda_{\text{QCD}}^4\left(G/\rho \right)^{1/2}\sim 10^6 \EV$, and so the favoured range of mass values $m_a\sim 10^{-6}-10^{-3}\mathrm{eV}$ for the QCD axion is deep into the regime dictated by gravity. This corresponds to a spatial scale $\sim 10^{12}\mathrm{m} \sim 10^{-4} \mathrm{lyr}$\cite{Guth:2014hsa}, which is comparable with the size of our solar system. That we are in this regime is evidence that the gravitational dynamics of the condensate is most important for galactic densities. We will therefore focus on the gravitational coupling for the remainder of the article. It would also be interesting to consider DS in the dense regime where the $\lambda$ term dominates. 
\subsection{Form of the clumps}
\label{sec:FormClumps}
Localized solutions of the axion field equations have been know for several decades~\cite{Jetzer:1991jr, Kolb:1993hw}, and were studied more recently in Refs.~\cite{Visinelli:2017ooc, Schiappacasse:2017ham, Hertzberg:2018lmt,Braaten:2018nag}. In this subsection we briefly review the facts we need from that literature. Using a spherically symmetric ansatz $\psi = \psi_0(r)\e^{-\ii \mu t}$ the equations of motion of Eq.~\eqref{eq:eom} (with no photon sources) reduce to a single dependent variable, and we can integrate them numerically using a shooting algorithm. The numerical profile is shown in Figs.~\ref{fig:approxphipsi}.

\begin{figure}
    \centering
    \includegraphics[width=0.5\textwidth]{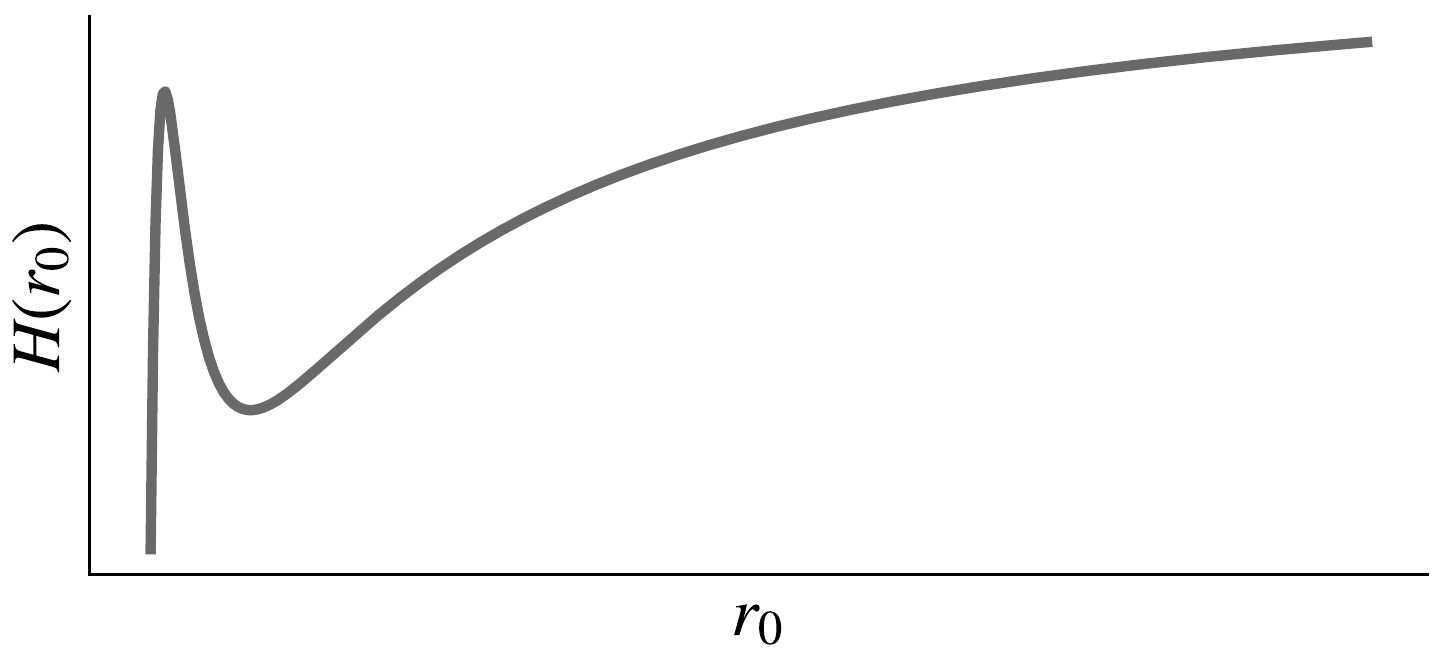}
    \caption{The effective potential of Eq.~\eqref{clumphamiltonian} as a function of the clump radius $r_0$. }
    \label{fig:veff}
\end{figure}

We can understand the dependence of the clump profile on the parameters in the problem by making the simple variational ansatz
\begin{align}
    \label{eq:expprofiles1}
    \psi_0(r) &= \sqrt{\frac{N}{\pi r_0^3}}\e^{-r/r_0}, \\
    \label{eq:expprofiles2}
    \Phi_0(r) &= -\frac{N G m^2}{r} \left( 1 - \e^{-2r/r_0} \right).
\end{align}
We can break up the radial coordinate into two regions: 
$\psi_0(r)=\sqrt{\frac{N}{\pi r_0^3}}\e^{-r/r_0}$ for an $N$-axion clump, where the radius $r_0$ is a variational parameter. The Hamiltonian evaluated on this ansatz becomes
\begin{align}
    \label{clumphamiltonian}
    H(r_0)\, &\equiv \int\dd^3 x\left[\frac{1}{2m}\lvert \nabla \psi\rvert^2 + \frac{\lambda}{16 m^2}\lvert\psi\rvert^4 - \frac{1}{2} G m^2 \int \dd^3 x' \frac{\lvert\psi({\bf x})\rvert^2 \lvert \psi({\bf x}')\rvert^2}{\lvert {\bf x}-{\bf x}' \rvert}\right]\nonumber\\
    &=\, \frac{1}{2m}\frac{N}{r_0^2} - \frac{5}{16}Gm^2\frac{N^2}{r_0} + \frac{1}{128\pi} \frac{\lambda}{m^2}\frac{N^2}{r_0^3}.
\end{align}
This is plotted in Fig.~\ref{fig:veff} (recall that $\lambda$ is negative). 
The idea is that when we minimize $H(r_0)$ with respect to $r_0$, the corresponding $\psi$ will approximate a clump solution with $\mu=H(r_0)/N$. 

For $N^2 \leqslant \frac{512\pi}{15 G\lvert \lambda\rvert m^2}$ Eq.~\eqref{clumphamiltonian} has two extrema: and unstable one at small radius, and a stable one at larger radius. If $N$ reaches this critical value, the saddles merge and there is no stable clump. Guided by the results of Sec.~\ref{sec:inhomogeneous}, we will consider the case where gravity is dominant, and so set $\lambda=0$. In this case there is a single stable clump solution corresponding to the minimum in Fig. \ref{fig:veff}, with the following parameters:
\begin{align}
    \label{eq:expresults}
    r_0 =\frac{16}{5}\frac{1}{(Gm^2)m N}, \qquad \mu \equiv \frac{H}{N} = -\frac{1}{2} \left(\frac{5}{16}\right)^2 (Gm^2)^2 m N^2,\\
    H_{\text{grav}} = - \left(\frac{5}{16}\right)^2 (Gm^2)^2 m N^2, \qquad H_{\text{kin}} = \frac{1}{2} \left(\frac{5}{16}\right)^2 (Gm^2)^2 m N^2. 
\end{align}
Notice that $(\text{Potential energy})=-2\times (\text{Kinetic energy})$, as expected from the virial theorem. 
\subsection{Dark sound normal modes}
\label{sec:SoundClumps}
\begin{figure}[t!bh]
	\centering
	\subfigure[]{\includegraphics[width=0.45\linewidth]{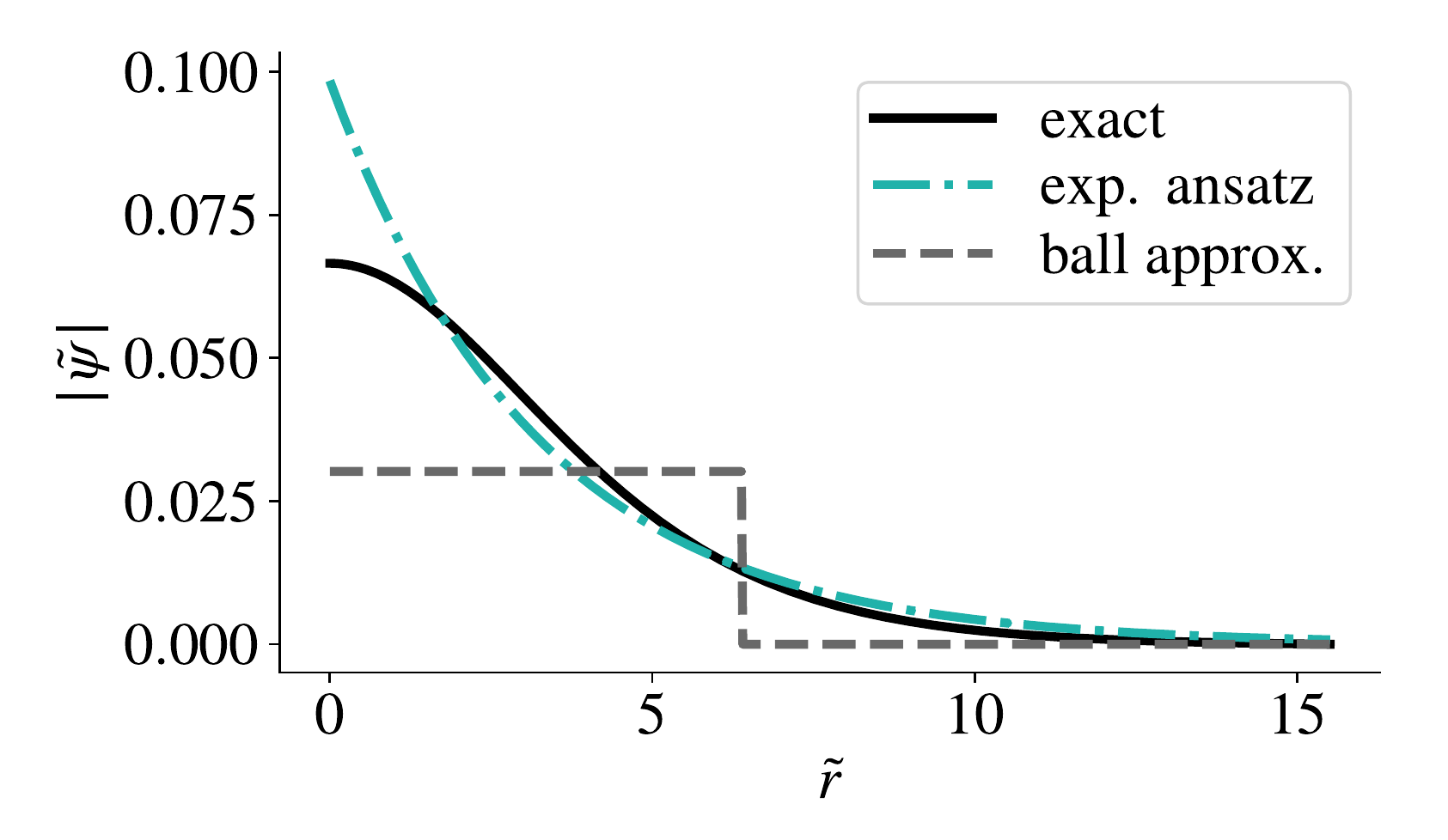} } \quad
	\subfigure[]{\includegraphics[width=0.45\linewidth]{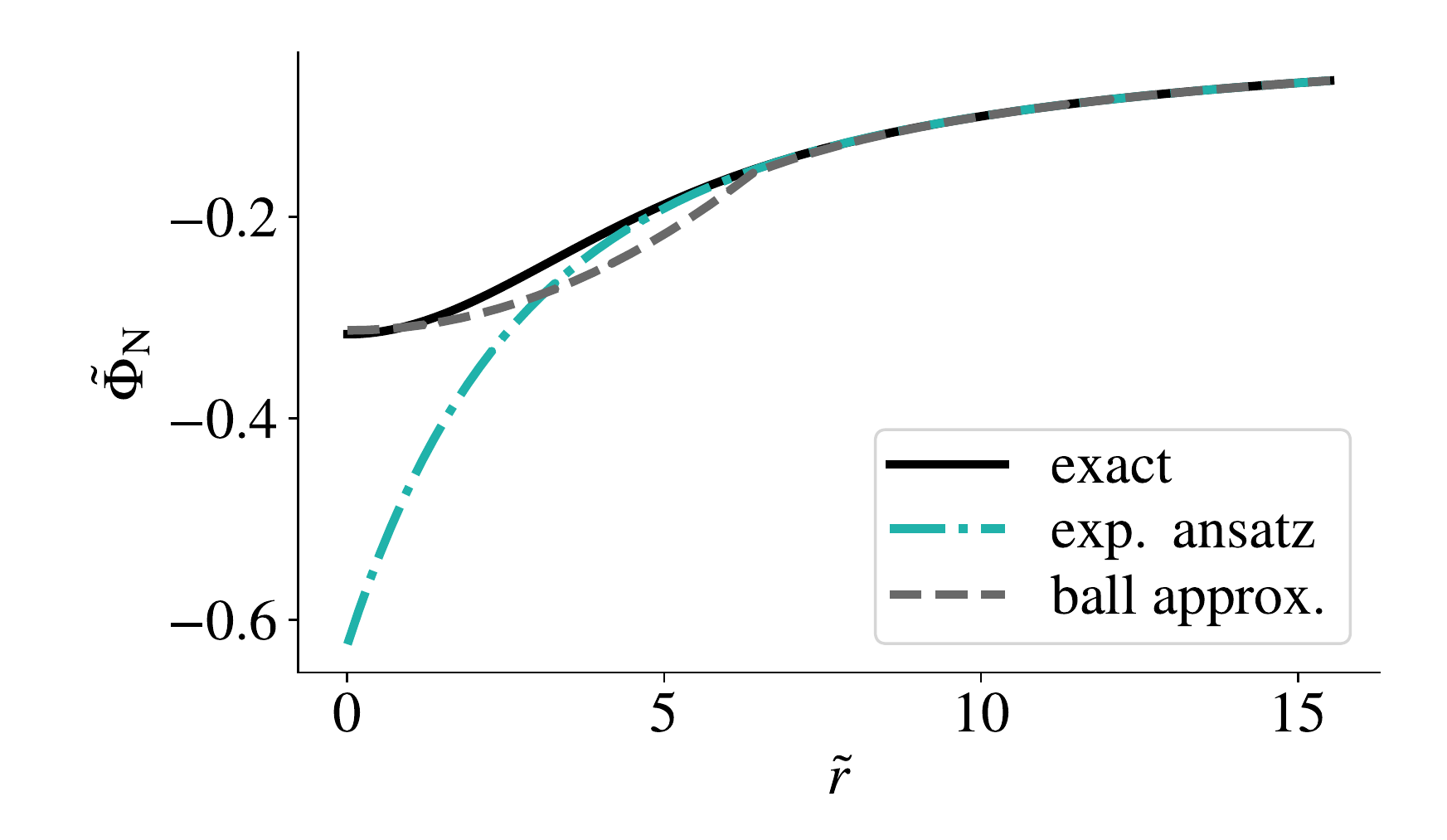}}
	\caption{(a) The axion profile $\tilde{\psi}$ and (b) the gravitational potential $\tilde{\Phi}_{\mathrm{N}}$ for the clump with $\tilde{N}=1$ (in dimensionless units defined in Sec.~\ref{sec:NumericalSimulations}). Black solid lines (`exact'): the numerically exact axion clump profile and potential found using a shooting algorithm. Cyan dot-dashed lines (`exp.~ansatz'): the exponential ansatz from Eqs.~\eqref{eq:expprofiles1} and \eqref{eq:expprofiles2}. Dashed lines (`ball approx.'): the uniform ball approximation from Eqs.~\eqref{eq:clumpapprox1} and \eqref{eq:clumpapprox2}.}
	\label{fig:approxphipsi}
\end{figure}
Because the clumps are the endpoint of the instability, the collective modes around them are stable. The finite size of the clumps means that the unstable modes around the homogeneous state are replaced by a discrete series of normal modes. They can be found by linearizing the equations of motion around the clump solution $\psi_0(r)$ in the manner of the classic Bogoliubov sound mode analysis~\cite{pitaevskii2003}. We use the ans\"{a}tze
\begin{align}
    \psi &= \e^{-\ii \mu t}\left[\psi_0 + \left(\alpha \e^{-\ii \omega t} + \beta \e^{\ii \omega t} \right)\right],  \label{eq:soundansatz1} \\
    \Phi_{\mathrm{N}} &= \Phi_0 + \Phi_1 \e^{-\ii \omega t} + (\Phi_1)^* \e^{\ii \omega t},
    \label{eq:soundansatz2}
\end{align}
where $\alpha$ and $\beta$ are functions of space having the interpretation of particles and holes in the condensate respectively. The linearized equations are (note that $\psi_0$ is real):
\begin{align}
    \label{eq:clumpbogoliubov1}
    (\mu + \omega)\alpha &= -\frac{1}{2m}\nabla^2\alpha + \frac{\lambda}{8m^2}\psi_0^2( 2\alpha + \beta^* ) + \Phi_0\alpha + \psi_0\Phi_1, \\
    \label{eq:clumpbogoliubov2}
    (\mu - \omega)\beta &= -\frac{1}{2m}\nabla^2\beta + \frac{\lambda}{8m^2}\psi_0^2( 2\beta + \alpha^* ) + \Phi_0\beta + \psi_0(\Phi_1)^*, \\
    \label{eq:clumpbogoliubov3}
    \nabla^2\Phi_1 &= Gm^2 \psi_0 (\alpha + \beta^*).
\end{align}
They differ from the standard Bogoliubov equations in having an extra component (the Newtonian potential) and spatially varying coefficients: they can be compared with the corresponding equations in e.g.~chapter 5 of~\cite{pitaevskii2003}. As before, we will set $\lambda$ to zero in order to concentrate on gravity. 

The energies of the Eqs.~\eqref{eq:clumpbogoliubov1}-\eqref{eq:clumpbogoliubov3} system should in principle be possible to obtain numerically by a shooting technique, but the appropriate boundary conditions are not clear and we have not been able to make this work. 

However, we can make progress by using an (albeit crude) approximation. Start with the exponential ansatz, which we repeat here for convenience of the reader (see Fig.~\ref{fig:approxphipsi})
\begin{align}
    \psi_0(r) &= \sqrt{\frac{N}{\pi r_0^3}}\e^{-r/r_0}, \\
    \Phi_0(r) &= -\frac{N G m^2}{r} \left( 1 - \e^{-2r/r_0} \right).
\end{align}
We can break up the radial coordinate into two regions: 
\begin{enumerate}
    \item $r\gtrsim 2r_0$: we are outside the clump and feel the $1/r$ gravitational potential of the spherically symmetric
    mass distribution.
    \item $r\lesssim 2r_0$: we are in the vicinity the clump, and feel the changing profile of the axion field. The $1/r$ behaviour of $\Phi_0(r)$ is cut off at this radius and instead approaches a finite value as $r\rightarrow 0$. 
\end{enumerate}
Therefore we model $\psi_0$ as a Heaviside step function of radius $2r_0$:
\begin{align}
    \label{eq:clumpapprox1}
    \psi_0(r) &\sim \sqrt{\frac{3N}{4\pi (2r_0)^3}}\, \Theta(2r_0-r), \\
    \label{eq:clumpapprox2}
    \Phi_0(r) &\sim N\, G m^2\times
    \begin{cases}
    \frac{1}{(2r_0)^3}(r^2 - (2r_0)^2) - \frac{1}{2r_0}  & r<2r_0\\
    -\frac{1}{r} & r>2r_0
    \end{cases},
\end{align}
for which the total gravitational energy is $H_{\mathrm{grav}}=-\frac{9}{15}\frac{N Gm^2}{2r_0}$, and using the virial theorem the chemical potential is $\mu = \frac{1}{2}H_{\mathrm{grav}}=-\frac{9}{30}\frac{N Gm^2}{2r_0}$. These approximations are shown in Fig.~\ref{fig:approxphipsi}. 

Even with this simplified ansatz, Eqs.~\eqref{eq:clumpbogoliubov1}-\eqref{eq:clumpbogoliubov3} are not simple to solve. They share many features with equations appearing in the theory of stellar pulsations in the Newtonian limit. In that case it is well-known practice to ignore the $\Phi_0$ to zeroth order, a step which is known as the Cowling approximation~\cite{Cowling1941}. We adopt this approximation. 

The equations for $\alpha$ and $\beta$ now decouple, and are each equal (up to a constant) to a simple system: the Schr\"{o}dinger equation for a harmonic oscillator,
\begin{align}
    (\mu+\omega)\alpha\, =\, -\frac{1}{2m}\nabla^2\alpha + \left(\frac{1}{2}m\Omega^2 r^2 + V_0 \right)\alpha, \\
    (\mu-\omega)\beta\, =\, -\frac{1}{2m}\nabla^2\beta + \left(\frac{1}{2}m\Omega^2 r^2 + V_0 \right)\beta,
\end{align}
where $V_0 = -\frac{Gm^2 N}{2r_0}$ and the fundamental oscillator frequency $\Omega = \sqrt{GmN/(2r_0)^3}$.  The predictions of our model end up being controlled by two numerical factors, which we call $\eta$ and $\gamma$. They are the proportionality constants in the relations:
\begin{align}
    r_0 &= \frac{\eta}{(Gm^2) m N}, \\
    \mu &= -\gamma (Gm^2)^2 m N^2.
\end{align}
As we saw above in Eq.~\eqref{eq:expresults}, the exponential ansatz predicts $\eta_{\mathrm{exp}} = 16/5$, $\gamma_{\mathrm{exp}}=\frac{1}{2} (\frac{5}{16})^2$, however in the next section we adjust these numbers with information from numerical simulations. In terms of $\eta$ and $\gamma$ the energy levels are
\begin{align}
    \mu \pm \omega = \frac{1}{\gamma}\left((\frac{3}{2}+n_x+n_y+n_z)\frac{1}{(2\eta)^{3/2}} - \frac{1}{2\eta}\right)\mu,
\end{align}
with $n_i$ being the occupation numbers each spatial direction. The value of $\omega$ for the ground state is $\pm\omega_{\mathrm{GS}}$, where the positive normal mode frequency $\omega_{\mathrm{GS}}$ has the expression
\begin{align}
\label{eq:groundstate}
    \omega_{\mathrm{GS}} = \left[\frac{1}{\gamma}\left(\frac{3}{2}\frac{1}{(2\eta)^{3/2}} - \frac{1}{2\eta}\right)-1\right]\mu\quad .
\end{align}
The salient feature of Eq.~\eqref{eq:groundstate} is that it predicts
\begin{align}
\label{eq:clumpnormalmode}
\boxed{\omega_{\mathrm{GS}} = b\, \mu}
\end{align}
with the numerical factor $b$ being universal for all clumps: $b = \left[\frac{3}{2}(2\eta)^{-3/2} - (2\eta)^{-1} \right]/\gamma-1$. In the next section, with aid from numerical simulations, we will estimate the number $b$ to be $\simeq 0.54$. The prediction of universality is only expected to be as good as the Cowling approximation we used to neglect the $\Phi_1$ terms in Eqs.~\eqref{eq:clumpbogoliubov1} and \eqref{eq:clumpbogoliubov2}, but as we will see it is still quite accurate. 

The condition that a state of the oscillator is bound to the clump is that the solution to the Schr\"{o}dinger problem has negative energy, which means that: $\mu-\omega < \frac{NGm^2}{2r_0}$. Using the same values of $\eta$ and $\gamma$ as before, this becomes $\frac{3}{2} + n_x + n_y + n_z \lesssim 2.53 $. This only definitely satisfied the ground state (the first excited state with a single $n_i=1$ also is below the bound but is very near the limit, so we cannot reach a definite conclusion in that case, given our approximations). Therefore we have the following prediction: a gravitationally bound axion clump has only one oscillation mode, which is spherically symmetric. This mode is the discretized dark sound of a gravitationally bound axion clump, when couplings to other field are absent. 
%
\subsection{Numerical simulations}
\label{sec:NumericalSimulations}
%
To gain more insight into the solution we performed numerical simulations of the clumps of Sec.~\ref{sec:SoundClumps} on a $100^3$ grid, using a pseudo-spectral method similar to the those in Ref.~\cite{Edwards:2018ccc}. At the edges of the grid we implemented a sponge potential as in Ref.~\cite{Hertzberg:2020dbk} to absorb outgoing radiation. We worked with rescaled units $\tilde{x}^i\equiv mx^i$, $\tilde{t}\equiv m t$, and $\tilde{\psi}\equiv (\sqrt{G m^2}/m^{3/2})\psi$, and correspondingly we define a rescaled potential and axion number $\tilde{\Phi}_{\mathrm{N}}\equiv (Gm^2/m) \Phi_{\mathrm{N}}$, $\tilde{N}\equiv Gm^2\, N$. 

\begin{figure}[htb]
    \centering
    \includegraphics[width=0.7\textwidth]{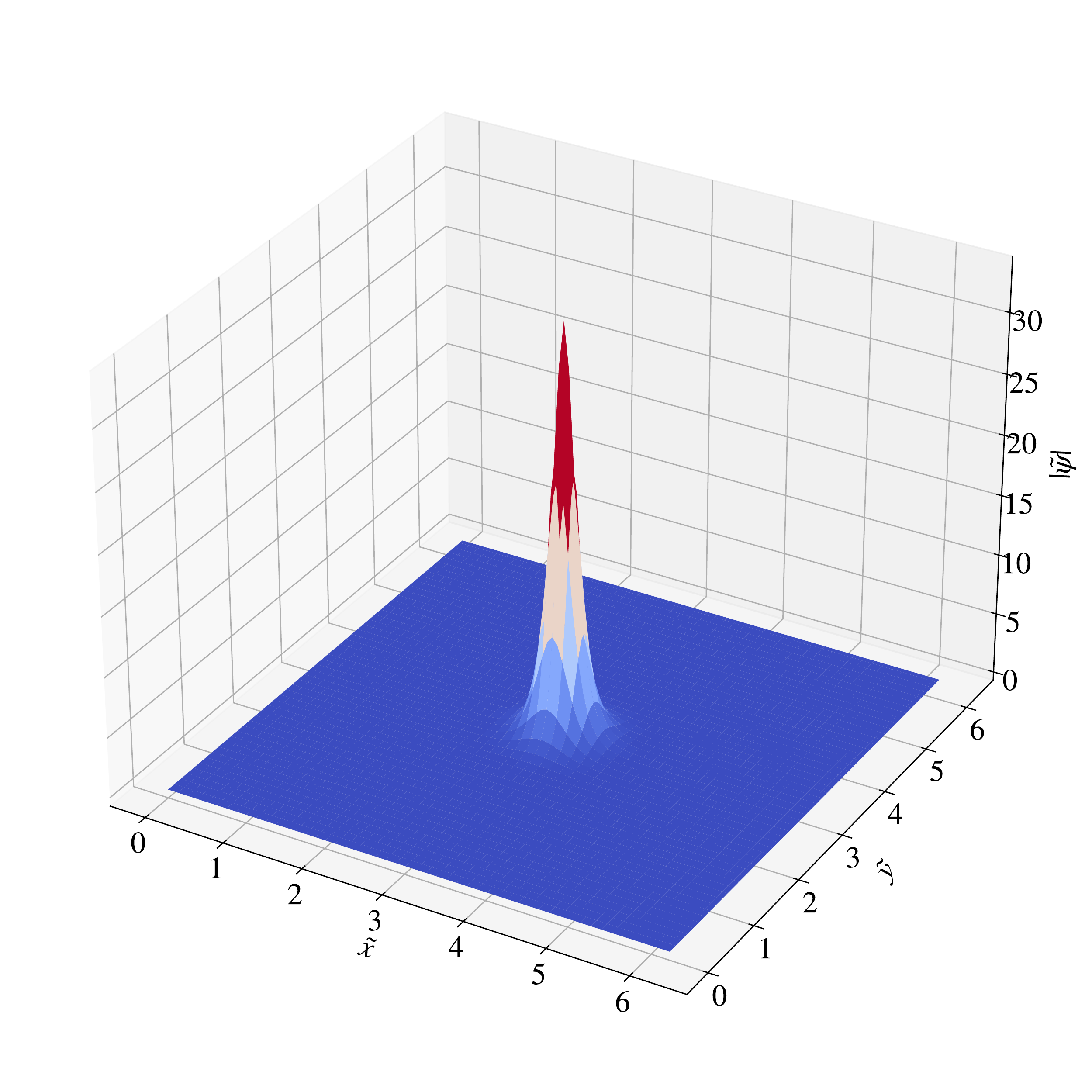}
    \caption{The absolute value of the (rescaled) field over a cross section through the centre of an oscillating clump. The peak field value oscillated with the frequencies in Fig.~\ref{fig:clumpnumerics}. }
    \label{fig:clumpimage}
\end{figure}

\begin{figure}[t!bh]
	\centering
	\subfigure[]{\includegraphics[width=0.45\linewidth]{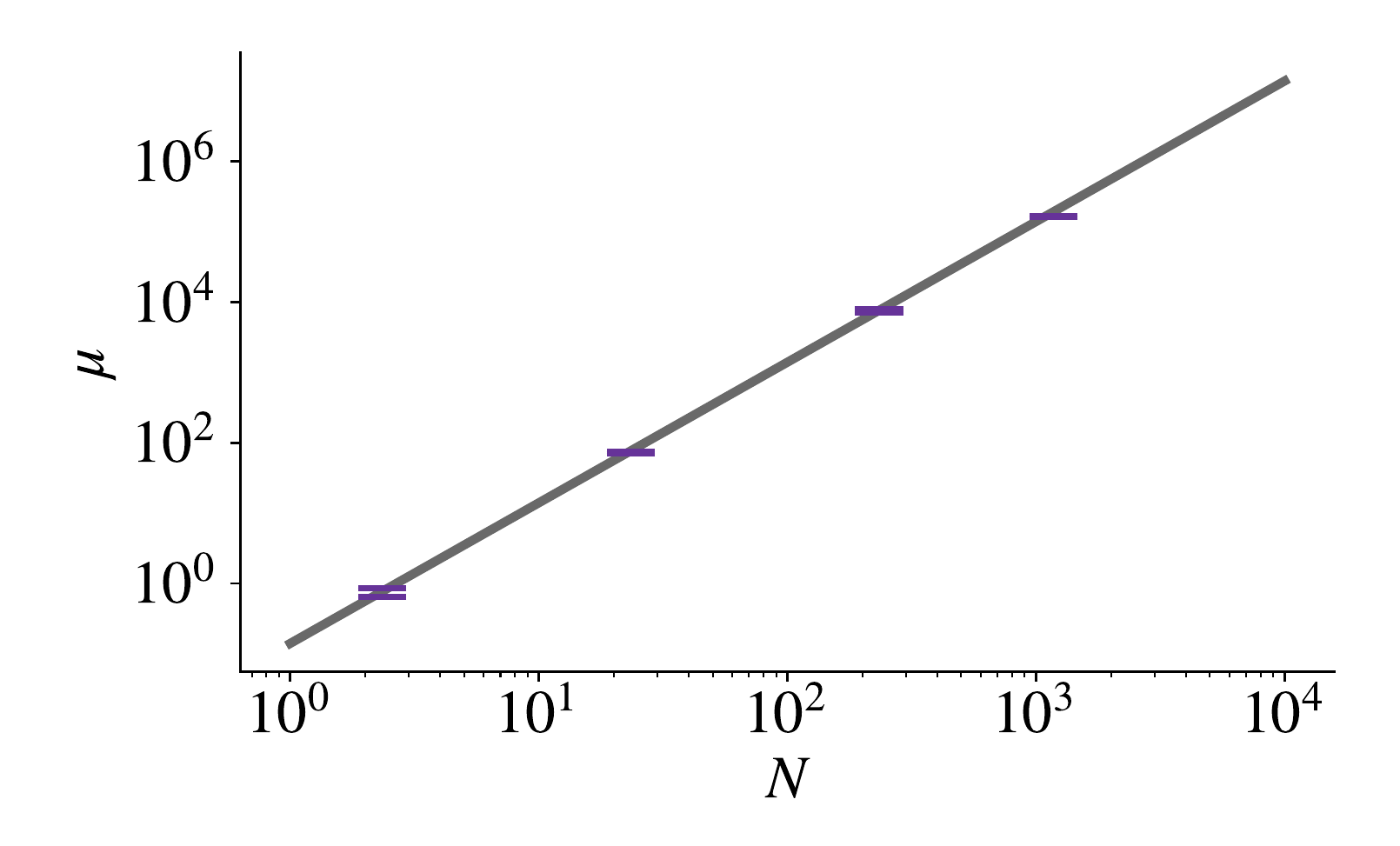} } \quad
	\subfigure[]{\includegraphics[width=0.45\linewidth]{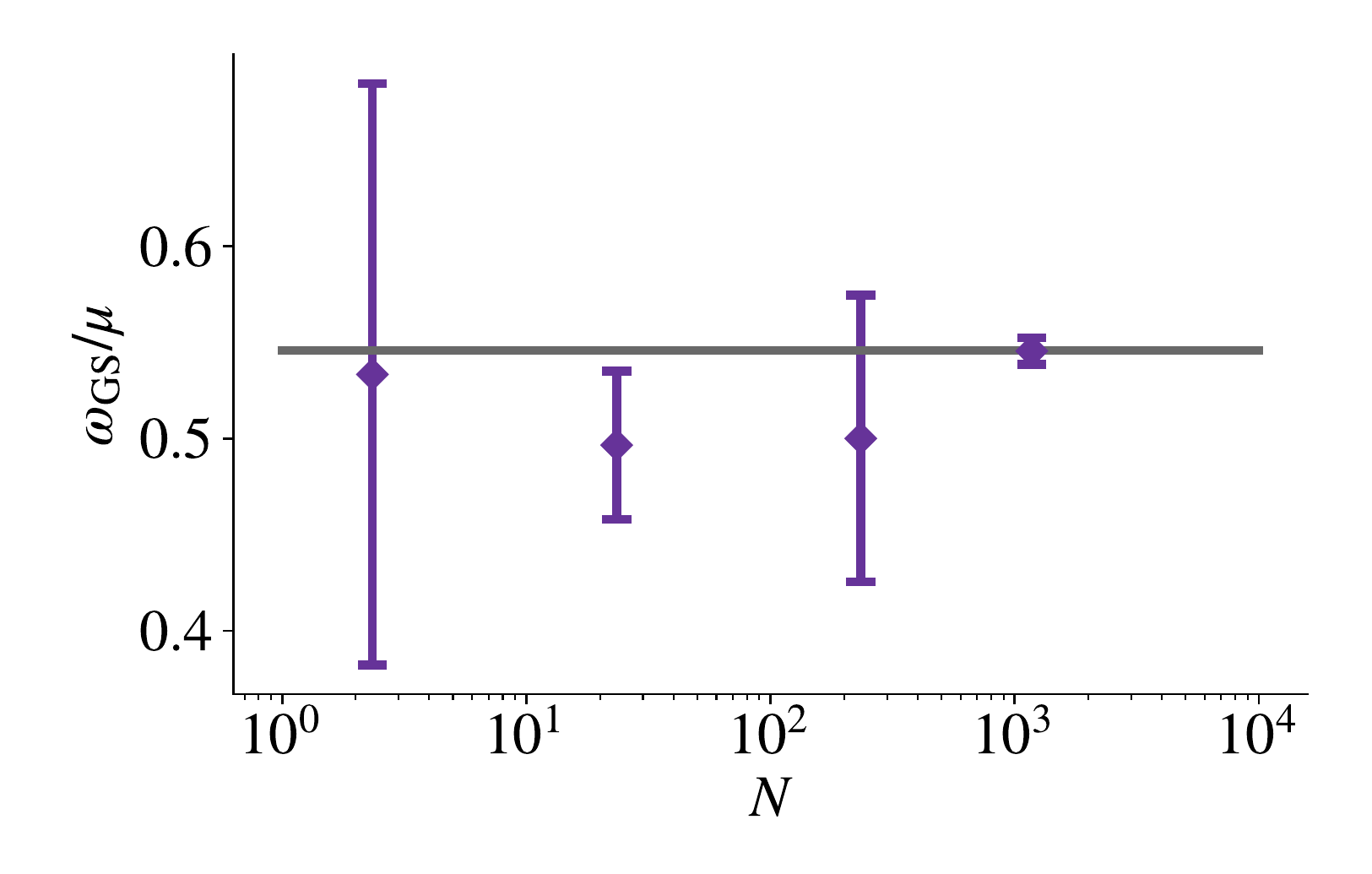}}
	\caption{Results (purple) from numerical simulations on a $100^3$ grid. Panel (a): the chemical potential $\mu$ of the clump as a function of number of axions in the clump $N$ on a log-log scale. Panel (b): the ratio of the oscillation frequency $\omega_{\mathrm{GS}}$ to the chemical potential. The full grey line shows the universal (semi-analytic) result of Eq.~\eqref{eq:GSomg}. The error bars come quantify error in estimating the centre value of finite-width peaks in plots such as Fig.~\ref{fig:clumpnumerics}. }
	\label{fig:numerics}
\end{figure}

\begin{figure}[t!bh]
\centering
    \includegraphics[width=0.5\textwidth]{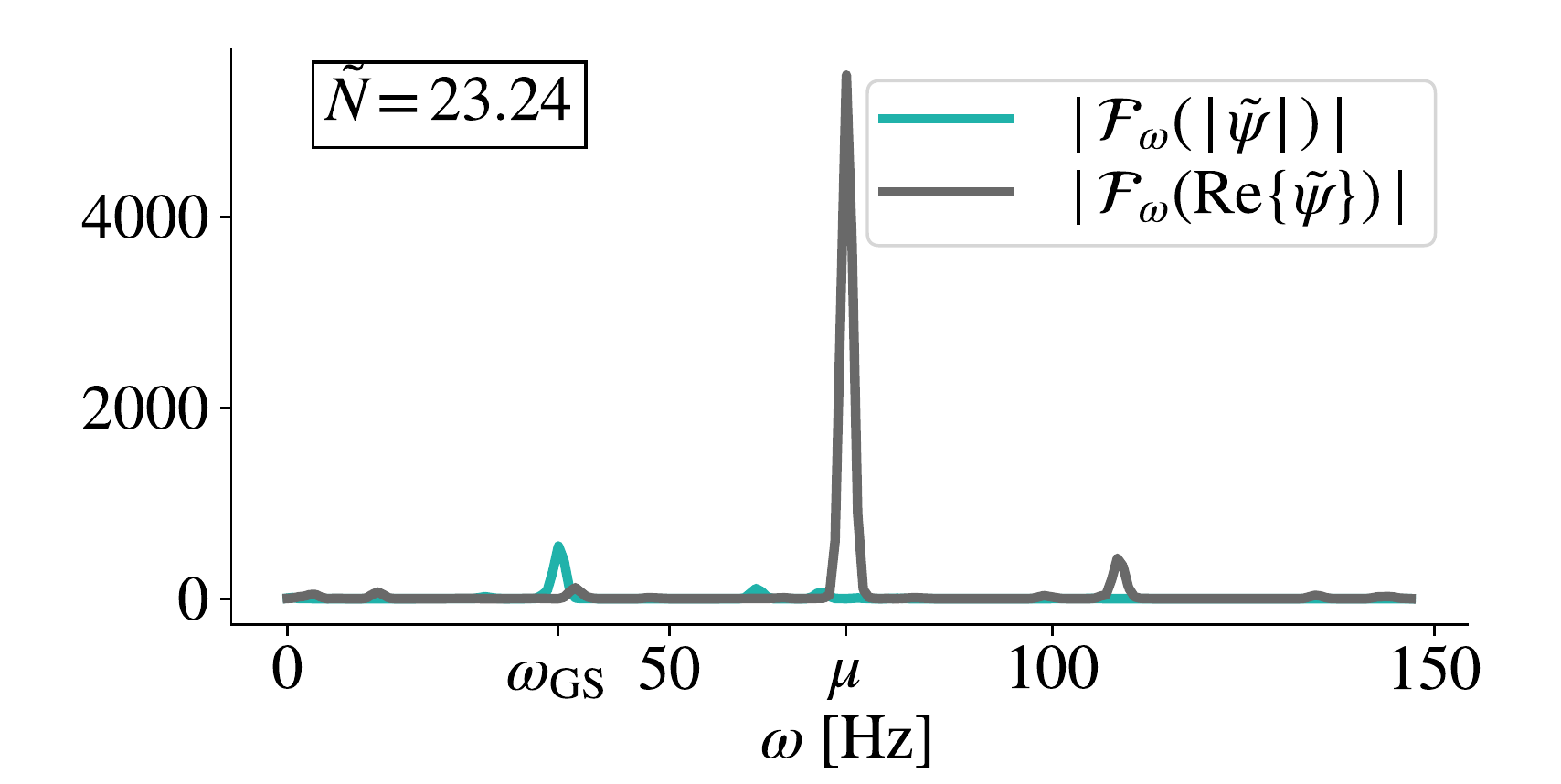}
    \caption{Frequency spectrum of the oscillation of an axion clump. The dominant peak in the spectrum of the magnitude, $\lvert \tilde{\psi} \rvert$ (cyan), is the ground state frequency $\omega_{\text{GS}}$, and the dominant peak in the real part, $\text{Re}\lbrace \tilde{\psi} \rbrace$ (grey), is the chemical potential.}
    \label{fig:clumpnumerics}
\end{figure}

An image of a clump in our simulations is shown in Fig.~\ref{fig:clumpimage}, and examples of data collected are appear in Figs.~\ref{fig:numerics} and \ref{fig:clumpnumerics}. We worked in units where the clump size in rescaled coordinates was of order one, but using the scaling symmetry under $x^i\rightarrow \lambda x^i$, $t\rightarrow \lambda^2 t$, $ \psi\rightarrow \lambda^{-1}\psi$ we can apply the results to a clump of any size.
We started the system out in a configuration slightly perturbed away from a stationary clump profile, and tracked the solution as it oscillated with its normal mode of Eq.~\eqref{eq:clumpnormalmode}. The values of $\mu$ observed in our numerical simulations are plotted in Fig.~\ref{fig:numerics}. We find that the value of $\gamma$ in Eq.~\eqref{eq:expresults} is larger than the observed value: the results in Fig.~\ref{fig:numerics} are instead fit well by $\gamma_{\mathrm{numerical}}\simeq 0.14$, which is the value mentioned in Sec.~\ref{sec:SoundClumps}. Using this value in Eq.~\eqref{eq:groundstate}, along with the $\eta=16/5$ from Eq.~\eqref{eq:expresults}, we obtain 
\begin{align}
    \label{eq:GSomg}
    \boxed{\omega_{\mathrm{GS}} \simeq \pm 0.54\mu}
\end{align}
This value is compared with numerics in Fig.~\ref{fig:numerics}. Notice the hybrid nature of our prediction: we used the analytic value for $\eta$ and combined it with the numerical value of $\gamma_{\text{numerical}}$ from simulations in order to arrive at \eqref{eq:GSomg}. 

Some details of our numerical simulations are as follows. We monitored the field values at the centre of an oscillating clump and made a frequency analysis, using the Blackman window function to mitigate spectral leakage, to obtain the results in Fig.~\ref{fig:clumpnumerics}. The large peaks in grey correspond to the chemical potential of the clump $\lvert \mu \rvert$. Equidistant on either side of this are smaller cyan peaks at $\omega\simeq\mu\pm 0.5\lvert \mu \rvert$, which are the normal modes. In absolute value the same peak occurs, consistent with our expectation from Eq.~\eqref{eq:soundansatz2}. 

A video from our numerical simulations is available to view at \url{https://tqmatter.org/dark-matter/} . 

%
\section{Observables and phenomenology}
\label{sec:observables}
%
In this section we provide numerical estimates for the DS parameters. Then we discuss possible experimental signatures, and the relation to the general picture of axion physics. 

%
\subsection{Estimates}
\label{sec:Estimates}
%
We use as benchmark a QCD axion with mass $1\mu \mathrm{eV}$ and density equal to the local dark matter density $\rho_{\mathrm{DM}} \sim 300\mathrm{MeV}\mathrm{cm}^{-3}$. We generally work in natural units, and use the expression $\rho\equiv m\lvert\psi\rvert^2$ for the axion density. These estimates are also summarized in Tab.~\ref{tab:EstimatesIntro}. 

The minimum sound speed of Eq.~\eqref{eq:soundspeed} is:
\begin{align}
v_s\, =\, \left(\frac{10^{-6}\EV}{m} \right)^{1/2}\left(\frac{\rho_{\text{DM}}}{\rho}\right)^{1/4} 10^{-12}c.
\end{align}
We estimate the gap produced by the $\boldsymbol{E}\cdot \boldsymbol{B}$ photon coupling using comparisons with the field strengths at the ADMX experiment ($B\sim 10\mathrm{T}$, $E\sim 10^6\mathrm{V}/\mathrm{m}$) \cite{Du:2020eej}:
\begin{align}
\label{eq:estimates}
\Delta &\sim \left(\frac{g_{a\gamma \gamma}}{10^{-15}\mathrm{GeV}^{-1}}\right)\left(\frac{\lvert \boldsymbol{E}\cdot \boldsymbol{B}\rvert}{10\mathrm{T}\cdot 10^6~\mathrm{Vm}^{-1}}\right)\left(\frac{\rho_{\text{DM}}}{\rho}\right)^{1/2} 10^{-18} \mathrm{eV},\\
\omega_{\mathrm{gravity}} &\sim \frac{k_s^2}{2m}\sim (G\rho)^{1/2} \sim \left(\frac{\rho}{\rho_{\text{DM}}}\right)^{1/2}10^{-30}\mathrm{eV}.
\end{align}
Next we turn to the clumps. The radius of a clump is
\begin{align}
    r_0\, \sim\, \frac{1}{(Gm^2 \rho)^{1/4}} \,\sim\, \left(\frac{10^{-6}\mathrm{eV}}{m} \right)^{1/2}\, \left(\frac{\rho_{\text{DM}}}{\rho}\right)^{1/4}\, 10^{12}\mathrm{m},
\end{align}
and the number of axions in this clump is
\begin{align}
    N\, \sim\, n\, r_0^3 \,\sim\, \left(\frac{10^{-6}\mathrm{eV}}{m} \right)^{5/2}\, \left(\frac{\rho}{\rho_{\text{DM}}}\right)^{1/4}\, 10^{56}.
\end{align}
The oscillation frequency $\mu$ of this clump is
\begin{align}
    \mu\, \sim\, (Gm^2)^2\, m N^2\, \sim\, \left(\frac{\rho}{\rho_{\text{DM}}}\right)^{1/2}\, 10^{-30}\mathrm{eV}.
\end{align}
\subsection{Shape of the axion signal}
\label{sec:SignalShape}
Understanding the nature of the axion signal in frequency space is important in interpreting experimental results. In terrestrial axion experiments this signal is commonly expected to have a line width dictated by the velocity dispersion of the local axion field~\cite{Budker:2013hfa}. In astrophysical searches, the signal shape may be determined by other factors. 

The DS interactions described in this paper modify the signal away from the expectation based on the free field. In the homogeneous case this is evident from Fig.~\ref{fig:Dispersion} where the IR dispersion relation is no longer quadratic. If we assume a thermal distribution of DS phonons, the expected distribution of axion frequencies is shown on the right hand side of Fig.~\ref{fig:Dispersion}. The gap $\Delta$ produces a clear peaked feature in the occupation number. The result in Eq.~\eqref{eq:estimates} corresponds to a frequency $\sim 10^{-3}\mathrm{Hz}$ for our benchmarks, which if detectable would be easier to see in the time domain. For astrophysical sources both the field strength and density could be much higher. For example, inside a neutron star field values could reach as high as $10^{11}\mathrm{T}$ \cite{Vidana:2018lqp} and $10^{21}\mathrm{V/m}$ \cite{Ray:2003gt}, giving a gap $\Delta\sim 10^{22}\mathrm{Hz}$. Therefore a broad range of signature frequencies is in principle possible. 

In the case of the clumps, we expect sidebands corresponding to the normal mode frequencies of the clump, as in Fig.~\ref{fig:clumpnumerics}. The scale of these features (again given in the last section) is $\sim 10^{-15}\mathrm{Hz}$ for galactic DM densities, corresponding to a time-scale of $\sim 10^{15}\mathrm{s}$, which is of cosmological scale ($\sim 100\mathrm{Myr}$). However the frequency increases with $\rho^{1/2}$. Therefore high density astrophysical environments could give rise to much greater frequencies. The dark matter density is often modelled with profiles that behave as $\sim 1/r$ or greater in the centre of galaxies~\cite{Linden:2014sra}. Some estimates put the DM density near our own galactic centre as high as $\sim 10^{11}\mathrm{GeV/cm}^3$~\cite{deLavallaz:2010wp}. Such densities give a timescale $1/\omega_{\mathrm{GS}}\sim 1\mathrm{yr}$, producing a variation in signal which it might be possible to detect in DM detection experiments. 

\subsection{Localized clump modes and specific heat}
\label{sec:HeatCapacity}
\begin{figure}[ht]
    \centering
    \includegraphics[width=0.5\textwidth]{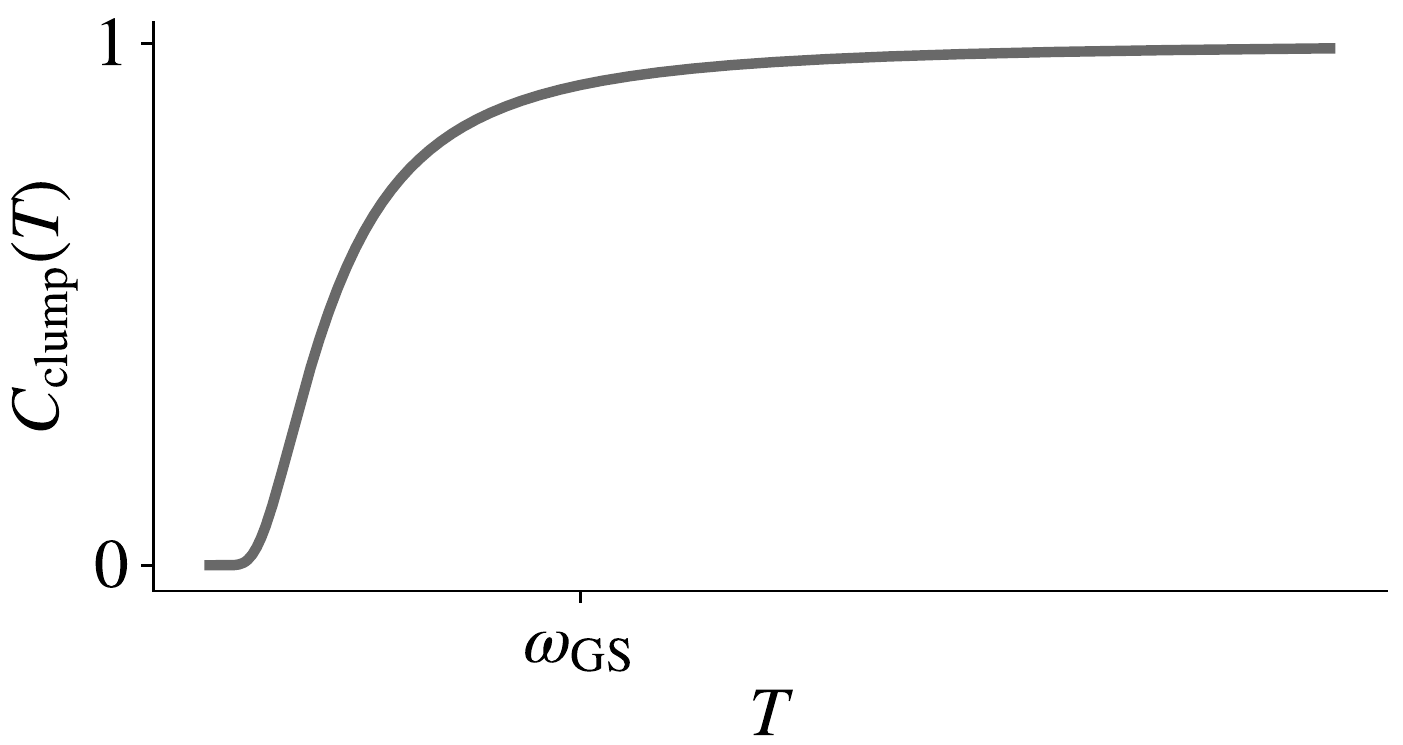}
    \caption{The heat capacity of an axion dark matter clump from Eq.~\eqref{eq:Cclump} as a function of temperature.}
    \label{fig:heatcapacity}
\end{figure}

Because the clump has a single DS mode, its heat capacity at equilibrium is that of a single harmonic oscillator of frequency $\omega_{\mathrm{GS}}$:
\begin{align}
    C_{\mathrm{clump}}\, =\,
    \label{eq:Cclump}
    \frac{(\frac{\omega_{\mathrm{GS}}}{T})^2}{4\sinh^2(\frac{\omega_{\mathrm{GS}}}{2T})}.
\end{align}
This is depicted in Fig.~\ref{fig:heatcapacity}. It jumps to finite values at $T\sim\omega_{\mathrm{GS}}$ corresponding to the energy of the first state (a single quantum of energy in the oscillation mode). As discussed in the last subsection, $\omega_{\mathrm{GS}}$ could vary between $10^{-15}\mathrm{eV}$ and $10^{-9}\mathrm{eV}$. On the other hand, given galactic DM velocity $\sim 10^{-3}c$ and using kinetic energy $\frac{1}{2} m v^2$, we can estimate the typical energy of axions to be $\sim 10^{-12}\mathrm{eV}$. It is therefore possible for significant energy to be stored in the normal modes of clumps in our galaxy, giving rise to the signal discussed above. 

For homogeneous DS coupled to photons, the heat capacity is similar to Fig.~\ref{fig:heatcapacity} but with $\Delta$ playing the role of $\omega_{\mathrm{GS}}$. Then the estimate in Eq.~\eqref{eq:estimates} of $\Delta\sim 10^{-18}\mathrm{eV}$ also allows DM modes to be occupied at galactic DM energies. 

We expect DS to produce fluctuations both in DM density and in photon occupation, similar to black body radiation of the EM field. Observation of these fluctuations could provide another route to DM observation. 
%
\section{Outlook and conclusion}
\label{sec:conclusion}
%
In this paper we discussed the axion DM density oscillations that result in DS modes. These modes are identical to longitudinal density oscillations, i.e.~the sound modes seen in any Bose liquid. For a homogeneous condensate we found that for wavelengths smaller than the instability size these are stable. 

Interaction with external fields, for example photons, gives rise to an approximate gap in the spectrum, and has the ability to stabilize the IR instability. We treated the photons as fixed external sources. When backreaction on the photons is considered it is possible that there exist stable structures of the coupled system, which will have a sound mode generalizing Eq.~\eqref{eq:dispersion}. 

Next we considered stable gravitationally bound axion clumps, and found a single sound normal mode at a frequency $\omega\simeq 0.54\mu$, where $\mu$ is the chemical potential of the clump. While preparing this paper we became aware of discussion of this mode in Ref.~\cite{Chiang:2021uvt}. 

We also discussed the implications of DS on observable quantities. Using estimates for dark matter densities at the centre of galaxies we found the possibility of variational effects on the timescale of $1\mathrm{yr}$ for clump normal modes. Typical DS frequencies can be greatly enhanced by coupling to photons: using field values in the ADMX experiment we found timescales of the order of hours ($\sim 10^{-3}\mathrm{Hz}$). 

Consequences of the DS modes will need to be explored. One can ask about the implications of the existence of DS in DM: 
\begin{itemize} \renewcommand{\labelitemi}{$\star$}
    \item How can we observe DS modes? Indeed it appears that DM is whispering into our ear. The situation is similar to the case of black body radiation in the universe where detectable radio noise due to thermal radiation is ubiquitous. The results of our work point to similar DS noise present in DM~\cite{Note5}. Perhaps one can come up with a detection scheme for this DS noise. 
    \item There are gravitational consequence of DS fluctuations.  If there are sound modes they represent the degrees of freedom that would be excited and hence induce DM density fluctuations down to the instability scale. It would be interesting to see what are the consequences of these inhomogeneities and what gravitational fingerprint they leave on the visible matter distribution. 
\end{itemize}
\begin{acknowledgments}
We are grateful to A.~J.~Millar, M.~Kopp, M.~Lawson, D.~Marsh, D.~J.~E.~Marsh, J. Conrad and F. Wilczek for useful discussions. 

This work was supported by the research environment grant `Detecting Axion Dark Matter In The Sky And In The Lab (AxionDM)' funded by the Swedish Research Council (VR) under Dnr 2019-02337. HSR and AVB were also supported by the University of Connecticut, the European Research Council under the European Unions Seventh Framework ERS-2018-SYG 810451 HERO and VILLUM  FONDEN via the Centre of Excellence for Dirac Materials (Grant No.~11744).
\end{acknowledgments}
\bibliographystyle{apsrev4-1}
\bibliography{dark_sound.bib}
\appendix
%
\section{Derivation of the non-relativistic action}
\label{sec:NonrelativisticAction}
%
Here we recap the derivation of the low energy action for the axion field in the presence of an EM field and small perturbations about the Minkowski metric. First write down the relativistic action Lagrangian for the axion field. As an example of an external coupling we consider a time-varying electromagnetic field at a single frequency, writing $N g_{a\gamma\gamma}\frac{e^2}{16\pi^2}\frac{1}{f_a}\mathbf{E}(t)\cdot\mathbf{B}(t) \equiv g\e^{\ii\omega_{\gamma}t}+\text{c.c.}$ to simplify notation. The Lagrangian is
\begin{align}
    \label{relaction}
    \mathcal{L}\, =\, \sqrt{-g}\left[\frac{1}{16\pi G}\mathcal{R}\, -\, \frac{1}{2}g^{\mu\nu}\pd_{\mu} a\pd_{\nu} a\, -\, \frac{1}{2}m^2 a^2\, -\, \frac{\lambda}{4!} a^4\+ (g\e^{\ii\omega_{\gamma}t}+\text{c.c.})\, a\+ \dots \right],
\end{align}
where the involved quantities are defined below Eq.~\eqref{eq:relaction}. We take the following limits:
\begin{enumerate}
    \item an approximately Minkowski metric: $g_{\mu\nu}=\eta_{\mu\nu} + h_{\mu\nu}$ with $h_{\mu\nu}\ll 1$ \label{lim1}
    \item a metric constant in time \label{lim2}
    \item non-relativistic axions: $\lvert \nabla a \rvert \ll m\, a$ \label{lim3}
\end{enumerate}
We can decompose the perturbation $h_{\mu\nu}$ of the metric into irreducible representations of the spatial rotations of the background Minkowski metric:
\begin{align}
    h_{00}\, &=\, -2\Phi_{\mathrm{N}},\\
    h_{0i}\, &=\, w_i,\\
    h_{ij}\, &=\, 2(\Sigma-\Phi_{\mathrm{N}}+S_{ij}),
\end{align}
corresponding to $\mathbf{10}\rightarrow \mathbf{1}+\mathbf{1}+\mathbf{3}+\mathbf{5}$ under $SO(3,1)\rightarrow SO(3)$. $\Phi_{\mathrm{N}}$ (the Newtonian potential) and $\Sigma$ are the two scalars, $w_i$ the vector, and $S_{ij}$ the traceless symmetric tensor. In limit \ref{lim1} the kinetic terms for these fields do not mix among themselves. 
In limit \ref{lim3}, $a$ couples to the metric only through the scalars. The lagrangian for these degrees of freedom becomes
\begin{align}
    \label{eq:newtonianaction}
    \mathcal{L}\, =\, &\frac{1}{2}\left[ \dot{a}^2 - (\nabla a)^2 - m^2 a^2 \, -\, \frac{\lambda}{4!} a^4\+ (g\e^{\ii\omega_{\gamma}t}+\text{c.c.})\, a\right]\\
    &+\, \frac{1}{8\pi G}\left[\Phi_{\mathrm{N}} \nabla^2 \Phi_{\mathrm{N}} - \Sigma \nabla^2 \Sigma \right]\, -\, \Phi_{\mathrm{N}} \left[2\dot{a}^2-m^2 a^2\right]\, +\, \frac{3}{2}\Sigma \left[\dot{a}^2-m^2 a^2\right]\quad .
\end{align}
In deriving \eqref{eq:newtonianaction} we have also neglected terms which are second order in the coupling constants $G,\lambda$ and $g$. 
Treating the field $a$ entirely classically (see the main text for a discussion of this assumption), we take the non-relativistic limit by inserting the mode expansion
\begin{align}
a\, \equiv \, \frac{1}{\sqrt{2m}}\left[ \psi(t,\mathbf{x})\, \e^{-\mathrm{i}(m+\mu)t} + \psi^{*}(t,\mathbf{x})\, \e^{\mathrm{i}(m+\mu)t} \right].
\end{align}
(the classical version of the quantum expression \eqref{eq:quantumfield}) into \eqref{eq:newtonianaction} and throwing away terms which oscillate fast, i.e. on the order of $\e^{-\mathrm{i}mt}$. In this limit we effectively set $\dot{a}^2=m^2 a^2$, and so the field combination coupling to $\Sigma$ vanishes. Thus $\Sigma$ decouples from the dynamics of interest. Integrating out the non-zero modes of $\Phi_{\mathrm{N}}$ we arrive at the action \eqref{nonrelaction}. The zero mode must be treated separately, and is not included the Newtonian description. It describes the cosmological expansion and will not be considered in this paper. 
%
\section{Momentum regime for homogeneous dark sound}
\label{sec:soundappendix}
%
We would like to know the range of momenta for which there is a well-defined sound mode. Thus we want to identify the range of $k$ values for which the dispersion relation of Eq.~\eqref{eq:dispersion} is approximately linear. Taylor expanding around $k_s$ (defined as the inflection point $\omega''(k_s)\equiv 0$), we have
\begin{align}
    \omega (k_s+\delta k)\, =\, \omega(k_s) + v_s \delta k + \frac{1}{3!}\omega'''(k_s)\, \delta k^3 + \cdots\quad .
\end{align}
The third derivative term will be smaller than 5\% of the first derivative term as long as
\begin{align}
    \delta k^2\, <\, 0.05 \times 3! \frac{v_s}{\omega'''(k_s)} \quad .
\end{align}
Thus we have a range of momenta for the sound regime:
\begin{align}
    \{k\}_{\mathrm{sound}}:\quad  k \in \left\{ k-\delta k, k+\delta k \right\},
\end{align}
where in the gravity case the values are given by
\begin{align}
    k_s= \left(48 \pi G m^2 \rho\right)^{1/4},\qquad\delta k<\sqrt{\frac{1}{20 \times 6^{1/2}}}\, k_s\simeq 0.14 k_s\quad ,
\end{align}
while for the self-interaction DS we have
\begin{align}
    k_s = \sqrt{\frac{3\lambda \rho}{4m^2}},\qquad \delta k<\sqrt{\frac{1}{40}}\, k_s\simeq 0.16 k_s\quad .
\end{align}
%
\section{Hydrodynamic equations}
\label{sec:HydrodynamicEquations}
%
In this appendix we write the axion dynamics in the form of fluid equations in order to highlight the analogies to that case and make contact with other literature. We start with the equations of motion written in terms of $n$ and $\theta$, where $\psi = \sqrt{n}\e^{\ii \theta}$:
\begin{align}
    \dot{n}\, &=\, -\frac{1}{m}\nabla\cdot (n\nabla\theta),\\
    \dot{\theta}\, &=\, \frac{1}{2m}\frac{\nabla^2\sqrt{n}}{\sqrt{n}} - \frac{1}{2m}\nabla^2\theta - \frac{\lambda}{8m^2} n - G m^2 \int\dd^3 x'
    \frac{\lvert \psi(\mathbf{x}')\rvert^2}{\lvert \mathbf{x}-\mathbf{x}'\rvert},
\end{align}
now if we define $\rho \equiv \rho n$ and $\mathbf{u}\equiv(\nabla\theta)/m$, we can massage these equations into the form
\begin{align}
    \label{eq:axionfluid}
    \dot{\rho}\, &=\, \nabla\cdot (\rho \mathbf{u})\\
    \partial_t(\rho \mathbf{u}) + \partial_j ( (\partial_j\theta) \rho \nabla\theta) \, &=\, \frac{\rho}{2m^2} \nabla\left( \frac{\nabla^2\sqrt{\rho}}{\sqrt{\rho}} \right) -\frac{1}{2m^2}\frac{\lambda}{8m^2}\nabla(\rho^2)  -  \frac{G}{m}\rho \nabla\left(\int\dd^3 x'\frac{\rho(\mathbf{x}')}{\lvert \mathbf{x}-\mathbf{x}' \rvert}  \right).
\end{align}
These are the standard continuity and Navier--Stokes equations of non-relativistic hydrodynamics. The terms on the right hand side of the second equation have the interpretation of the minus the pressure gradient. The first term is often called the `quantum pressure', and is a manifestation of the Heisenberg uncertainty principle: localizing the particles to a spatial region will produce a uncertainty in their velocity, and corresponding pressure term. However it cannot be written in terms of the gradient of a local function, so is a sort of non-local pressure. The second term a standard collisional term coming from the point interactions of the axion. The third term is also non-local, since it comes from the Newtonian gravitational force. 

The crucial point to notice is that all the pressure terms have negative sign apart apart from the quantum pressure.  In standard treatments of the condensate sound mode~\cite{Stringari:1996zza} this term is neglected because it is subleading in spatial derivatives, and the $\lambda$ term produces a linear dispersion relation at low momentum. This is not possible in our case because the quantum pressure is the only repulsive force in the system. 

Linearizing Eq.~\eqref{eq:axionfluid} about the homogeneous state ($\rho(x) = \rho_0 + \delta\rho (x)$, $\lvert \mathbf{u}\rvert \ll 1$) and taking appropriate derivatives we obtain a closed equation for the density $\delta\rho$ which is a non-local generalization of the standard wave equation for sound:
\begin{align}
    \label{eq:soundequation}
    \partial_t^2 \delta\rho \, =\, &\frac{1}{2m^2}\partial_i \left[ \partial_i  \left(\frac{\nabla^2\sqrt{\rho}}{\rho}\right)  \delta\rho\, +\, \frac{1}{2} \rho \partial_i \left( \frac{\nabla^2(\delta\rho/\sqrt{\rho})}{\sqrt{\rho}} \right)
    \, -\, \frac{1}{2} \rho \partial_i \left( \frac{\delta\rho}{\rho^{3/2}}  \nabla^2\sqrt{\rho} \right) \right]\\
    &+\, \frac{1}{2m^2}\frac{\lambda}{8m^2}\nabla\left( \rho\nabla(\delta\rho) \right) -\, \frac{G}{m}\left[ \nabla\left( \delta\rho \nabla\int\dd^3 x'
    \frac{\rho(\mathbf{x}')}{\lvert \mathbf{x}-\mathbf{x}'\rvert}\right)\, +\, \nabla\left( \rho \nabla\int\dd^3 x'
    \frac{\delta\rho(\mathbf{x}')}{\lvert \mathbf{x}-\mathbf{x}'\rvert}\right) \right].
\end{align}
The fact noted above, that we cannot ignore quantum pressure, means that the top line of Eq.~\eqref{eq:soundequation} cannot be neglected and we must always solve a fourth order differential equation. This implies that an extra boundary condition is necessary compared to a second order wave equation. We have not been able to identify what it should physically correspond to. This would be interesting to consider in future work. 
\end{document}